\documentclass[aps,prapplied,reprint,longbibliography]{revtex4-2}
\usepackage{array}[=2016-10-06]
\usepackage{amsmath,amssymb}
\usepackage{graphicx}
\usepackage{booktabs}
\usepackage{dcolumn}
\usepackage{xcolor}
\usepackage{siunitx}
\usepackage{microtype}
\usepackage{hyperref}
\definecolor{linkblue}{rgb}{0.10,0.20,0.45}
\hypersetup{colorlinks=true,linkcolor=linkblue,citecolor=linkblue,urlcolor=linkblue,
  pdfauthor={},pdfcreator={},pdfproducer={}}
\DeclareSIUnit\angstromunit{\text{\AA}}

\newcommand{\Ek}{E_k}
\newcommand{\Eeff}[1][i]{E_{k,#1}^{\mathrm{eff}}}
\newcommand{\vlam}{\vec{\lambda}}
\newcommand{\wmin}{\omega_{\min}}
\newcommand{\fmin}{f_{\min}}
\newcommand{\vg}{v_g}
\newcommand{\vp}{v_p}
\newcommand{\meff}{m_{\mathrm{eff}}}
\newcommand{\ldecay}{\ell_{\mathrm{decay}}}
\newcommand{\kB}{k_B}
\newcommand{\Tr}{\operatorname{Tr}}
\newcommand{\diff}{\mathrm{d}}
\newcommand{\EnvI}{Envelope~I}
\newcommand{\EnvII}{Envelope~II}

\begin{document}

\title{Polarization waves in open-barrier quantum-dot cellular automata}

\author{Konrad Walus}
\email{konradw@ece.ubc.ca}
\affiliation{Department of Electrical and Computer Engineering, The University of British Columbia, Vancouver, British Columbia V6T 1Z4, Canada}

\date{November 13, 2026}

\begin{abstract}
Quantum-dot cellular automata (QCA) are almost always operated as bistable digital logic.
We develop a closed-form theory of small-signal polarization waves in the open-barrier regime, where tunnelling within each cell is comparable to or stronger than the electrostatic kink coupling between cells, so that a line of cells responds linearly instead of latching.
Linearizing the coherence-vector equations of the intercellular Hartree approximation (ICHA) about the unpolarized state gives a lattice dispersion with a low-frequency cutoff, an evanescent band below it and a Klein--Gordon long-wavelength limit, and relaxation toward the instantaneous equilibrium damps the waves at a rate that depends on the wavenumber.
The exact Jordan--Wigner solution and exact diagonalization confirm the mode as a sharp quasiparticle, with ICHA frequencies within \SI{6}{\percent} of the exact ones once the tunnelling energy is at least twice the kink energy, and show that the exact chain also carries waves over a range of couplings where the ICHA predicts latching.
A leading low-temperature estimate of thermal dephasing sets a ceiling on the operating temperature of about \SI{13.6}{K} per terahertz of gap for a carrier that keeps its phase over 30 periods.
At junctions the waves scatter through a lattice scattering matrix that reaches the familiar transmission-line value only at mid-band.
\end{abstract}

\maketitle

\section{Introduction}
\label{sec:intro}

Quantum-dot cellular automata (QCA) encode information in the charge configuration of coupled quantum dots: in each cell two configurations, with polarization $P=\pm1$, serve as binary states~\cite{lent1993quantum,tougaw1994logical}, and neighbouring cells interact electrostatically, so a line of cells copies a polarization from cell to cell.
QCA has been developed almost entirely as digital logic~\cite{lent1997device,walus2004qcadesigner}, in the high-barrier regime where the intracell tunnelling energy $\gamma$ is small compared with the intercell kink energy $\Ek$ and each cell latches into a saturated state; switching relies on adiabatic clocking~\cite{lent1997device}, whose power gain and dissipation were analysed by Timler and Lent~\cite{timler2002power}.
QCA cells and gates have been demonstrated with aluminium islands~\cite{orlov1997realization,amlani1999digital,orlov2000experimental} and phosphorus-implanted silicon dots~\cite{mitic2006demonstration}, molecular cells have been designed~\cite{lent2003molecular}, and atomic-scale cells built from silicon dangling bonds (SiDBs) realize binary wires and an OR gate~\cite{huff2018binary}.

Coupled quantum dots can also act as analog dynamical systems.
Quantum cellular neural networks treat arrays of coupled cells as analog processing elements~\cite{toth1996qcnn}.
Csurgay et al.\ formulated signal processing with near-neighbour-coupled, time-varying quantum-dot arrays in the coherence-vector picture and linearized a driven cell about its equilibrium to obtain its one-port small-signal admittance~\cite{csurgay2000signal}.
Fortuna et al.\ analysed the nonlinear dynamics of quantum-dot cells and one-dimensional arrays: chaotic oscillation and local-activity conditions for two coupled cells, and synchronization along arrays~\cite{fortuna2004nanoscale}.
Any analog use requires a quantitative account of how small signals propagate, disperse, decay and scatter in an array that does not latch.
This paper gives that account for the open-barrier regime, $\gamma\gtrsim\Ek$, in which the cell-to-cell response is soft and linear for small signals.

\paragraph*{Prior work.}
Lent and Tougaw introduced the intercellular Hartree approximation (ICHA) for the ground states of QCA arrays~\cite{lent1993lines,tougaw1994logical}, and Tougaw and Lent extended it to time-dependent dynamics, identifying the two-state QCA line as a transverse-field Ising chain and finding that intercellular correlations speed up kink propagation~\cite{tougaw1996dynamic}.
T\'oth and Lent wrote the ICHA in coherence-vector form, compared it with exact correlated dynamics during quasi-adiabatic switching, and constructed intermediate pair-correlation models~\cite{toth2001role,toth2000thesis}, which Karim and the present author extended to arbitrary circuits with relaxation~\cite{karim2014efficient}.
Taucer et al.~\cite{taucer2015consequences} found that the ICHA can predict a spurious latching of small groups of cells and that intercellular entanglement matters for molecular and atomic cells, and showed that relaxation brings isolated groups of cells to an unpolarized state.

Govorov and Chaplik studied collective modes and ferroelectric-like transitions of tunnel-coupled quantum-dot arrays~\cite{govorov1994ferroelectric,chaplik1996quantum}, and a transverse-field Ising chain of superconducting charge qubits has been proposed for observing nonequilibrium dynamics~\cite{viehmann2013observing}.
T\'oth~\cite[Sec.~4.2]{toth2000thesis} linearized the ICHA about the polarized (latched) state of a line with $\gamma<\Ek$, obtained the narrow band $\hbar^2\omega^2=4(\Ek^2-\gamma^2\cos ka)$, whose lower edge is the in-phase mode, and showed in simulation that a driven line transmits only inside this band and reflects at its far end; he also simulated large-amplitude wavefronts and their collisions~\cite[Sec.~4.3]{toth2000thesis}.
We linearize the same equations about the other equilibrium, the unpolarized state of the open-barrier regime.
The resulting dispersion is the random-phase (pseudo-spin-wave, or tunnelling-mode) dispersion of the transverse-field Ising model known from order--disorder ferroelectrics, where the soft mode and the comparison with the exact chain are classical~\cite{degennes1963collective,brout1966tunnelling,blinc1974soft,stinchcombe1973ising}; the chain is solved exactly by a Jordan--Wigner transformation~\cite{pfeuty1970one}; and at fixed frequency the junction problem of Sec.~\ref{sec:junctions} is the scattering at a vertex that joins semi-infinite tight-binding chains~\cite{datta1995electronic,texier2001scattering}.
Applying these established descriptions to small-signal transport in open-barrier QCA lines, this paper adds the damping, temperature and junction results and the quantitative benchmark against the exact chain.

In the saturated regime Nakagawa et al.\ obtained kink solitons that are transmitted or reflected at junctions~\cite{nakagawa2001kink,nakagawa2001junction}; in the two-state picture Tougaw and Lent had already described a single kink as a tight-binding pseudo-particle hopping between cells~\cite{tougaw1996dynamic}. Our results concern linear waves.
Unitary ``quantum cellular automata'', discrete-time models whose continuum limits give Dirac and Klein--Gordon equations~\cite{bialynickibirula1994weyl,meyer1996quantum}, belong to a different field; here QCA always means quantum-dot cellular automata.
The Klein--Gordon limit found below (Sec.~\ref{sec:continuum}) has a different origin: it is the long-wavelength form of the polarization-wave dispersion of a gapped line of cells, with the gap in the role of the mass.

\paragraph*{Clocked wires.}
In earlier work, Retallick and the present author~\cite{retallick2020limits} obtained the low-energy spectrum of driven QCA wires analytically with a modified Jordan--Wigner transformation: the minimum gap of an adiabatic clock cycle and the width of that avoided crossing both fall as $1/(N+1)$ for a wire of $N$ cells, so that in a Landau--Zener estimate the maximum adiabatic clock rate in the coherent limit falls as $(N+1)^{-2}$.
A clock drives the line through the critical point of the same chain, where our exact gap, $N\Delta\to(\pi/4)\Ek$ (Appendix~\ref{app:ed}), is consistent with that $1/N$ scaling; analog operation stays in the gapped paramagnetic phase, where the signal speed is set by the maximum group velocity $a\Ek/\hbar$ (Sec.~\ref{sec:exact}) rather than by critical slowing down.

\paragraph*{Results.}
Read as a signal medium, the open-barrier line is high-pass, with a cutoff, an evanescent band and a Klein--Gordon limit (Sec.~\ref{sec:ac}).
Relaxation toward the instantaneous equilibrium damps the waves at a wavenumber-dependent rate $\beta(k)$ (Sec.~\ref{sec:damping}), and junctions scatter them with a lattice matrix that reproduces the transmission-line picture only at mid-band (Sec.~\ref{sec:junctions}).
The exact solution (Sec.~\ref{sec:exact}) confirms the ICHA mode as a sharp quasiparticle for $\gamma\ge2\Ek$, replaces the ICHA dispersion by the exact quasiparticle dispersion through $\varepsilon_k^2=\hbar^2\omega_k^2+\Ek^2$, and identifies the asymptotic ballistic-front velocity with the known maximal quasiparticle velocity $a\Ek/\hbar$ of the chain~\cite{calabrese2011quench,calabrese2012quench}, rather than the $\gamma$-dependent ICHA value.
For $\Ek/2<\gamma<\Ek$ the exact chain carries waves where the ICHA predicts latching, a quantitative, small-signal instance of the spurious ICHA latching noted by Taucer et al.~\cite{taucer2015consequences}.
The leading stationary low-temperature estimate of thermal-quasiparticle dephasing in the nearest-neighbour transverse-field Ising chain ties the temperature ceiling to the gap, about \SI{13.6}{K} per terahertz of gap for a carrier that keeps its phase over $N_c=30$ periods, a criterion we choose (Sec.~\ref{sec:tmax}).
Within the stated relaxation model the damping rate $\beta(k)$ is exact (Sec.~\ref{sec:damping} and Appendix~\ref{app:derivations}).

\section{From the four-dot cell to the coherence-vector equations}
\label{sec:model}

This section first summarizes the standard QCA description in the conventions of T\'oth and Lent and then defines the regime and the small-signal problem studied here. Readers who know the coherence-vector description can skip to the paragraph on the open-barrier regime. The symbols are listed in Appendix~\ref{app:notation}.

\paragraph*{The four-dot cell.}
The canonical QCA cell has four quantum dots at the corners of a square and holds two excess electrons~\cite{lent1993quantum}.
Coulomb repulsion places the electrons on opposite corners, so the cell has two degenerate low-energy configurations, one on each diagonal.
With $Q_d$ the excess charge on dot $d$, numbered around the square, the polarization $P=[(Q_1+Q_3)-(Q_2+Q_4)]/(Q_1+Q_2+Q_3+Q_4)$ is $+1$ on one diagonal and $-1$ on the other~\cite{lent1993quantum}.
A pair of dots holding one electron, such as a pair of SiDBs~\cite{huff2018binary}, has the same two-state structure.

\paragraph*{The two-state approximation.}
Tunnelling between dots couples the two configurations; configurations with both electrons on one side of the cell cost an intracell Coulomb energy that is large compared with the tunnelling and intercell energies, and we drop them.
Tougaw and Lent tested a reduced basis against the full sixteen-dimensional site-state basis for two opposite-spin electrons on four dots: for their standard GaAs cell between equally polarized drivers, the ground-state projection onto their self-consistent, symmetrically orthogonalized two-state subspace remained at least 0.993 over the driver-polarization sweep~\cite{tougaw1996dynamic}.
Here we model each cell as a two-level system with basis states $|{+1}\rangle$ and $|{-1}\rangle$, the configurations with $P=\pm1$; $|{+1}\rangle$ is the first basis vector.
Following T\'oth and Lent~\cite{toth2001role}, who follow Mahler and Weberru{\ss}~\cite{mahler1998quantum}, we use the Pauli matrices
\begin{equation}
  \sigma_x=\begin{pmatrix}0&1\\1&0\end{pmatrix},\quad
  \sigma_y=\begin{pmatrix}0&\mathrm i\\-\mathrm i&0\end{pmatrix},\quad
  \sigma_z=\begin{pmatrix}-1&0\\0&1\end{pmatrix},
  \label{eq:pauli}
\end{equation}
in which $\sigma_y$ and $\sigma_z$ have the opposite sign to the usual convention, a rotation by $\pi$ about the $x$ axis that keeps the usual algebra, $[\sigma_a,\sigma_b]=2\mathrm i\,\epsilon_{abc}\sigma_c$.
The polarization operator is $|{+1}\rangle\langle{+1}|-|{-1}\rangle\langle{-1}|=-\sigma_z$, so
\begin{equation}
  P=-\langle\sigma_z\rangle .
  \label{eq:P_def}
\end{equation}

\paragraph*{The cell Hamiltonian and the kink energy.}
In the two-state basis~\cite{tougaw1996dynamic} the Hamiltonian of cell $i$, in the form used by T\'oth and Lent and by Timler and Lent~\cite{toth2001role,timler2002power}, reads
\begin{align}
  H_i&=\begin{pmatrix}-\frac{\Ek}{2}\sum_jP_j&-\gamma\\[2pt]-\gamma&+\frac{\Ek}{2}\sum_jP_j\end{pmatrix}\nonumber\\
  &=-\gamma\,\sigma_x+\frac{\Ek}{2}\Bigl(\sum_jP_j\Bigr)\sigma_z ,
  \label{eq:hamiltonian}
\end{align}
where the sum runs over the neighbours of cell $i$.
Here $\gamma$ is the tunnelling energy between the two configurations, and $\Ek$ is the kink energy: the electrostatic energy cost of two neighbouring cells with opposite polarizations relative to two with the same polarization.
A positively polarized neighbourhood lowers the energy of $|{+1}\rangle$, so neighbouring cells tend to align.
We write $\Eeff=\Ek\sum_jP_j$ for the field of the neighbours, or $\Eeff=\sum_jE_{k,ij}P_j$ when the couplings differ.
Then $H_i=\tfrac12\vec\Gamma_i\cdot\vec\sigma$, with
\begin{equation}
  \vec\Gamma_i=\bigl(-2\gamma,\;0,\;\Eeff\bigr),\qquad \Gamma_{i,a}=\Tr(\sigma_aH_i) .
  \label{eq:gamma_vec}
\end{equation}

\paragraph*{The nonlinear cell-to-cell response.}
For a cell driven by one neighbour of fixed polarization $P_{\mathrm{in}}$, the ground state of Eq.~\eqref{eq:hamiltonian} has the polarization (Appendix~\ref{app:derivations})
\begin{equation}
  P_{\mathrm{out}}=\frac{\Ek P_{\mathrm{in}}/2\gamma}{\sqrt{1+(\Ek P_{\mathrm{in}}/2\gamma)^2}} .
  \label{eq:response}
\end{equation}
For weak tunnelling, $\gamma\ll\Ek$, Eq.~\eqref{eq:response} is a steep, saturating step, the bistable response on which digital QCA rests~\cite{lent1993quantum,tougaw1994logical}: a small input polarizes the cell almost fully.
Its slope at $P_{\mathrm{in}}=0$, $\Ek/2\gamma$, is the small-signal gain of one cell-to-cell link~\cite{toth2000thesis}.

\paragraph*{Many cells, exactly and in the ICHA.}
For a line of cells with nearest-neighbour coupling the many-cell Hamiltonian is
\begin{equation}
  H=-\gamma\sum_i\sigma_x(i)-\frac{\Ek}{2}\sum_i\sigma_z(i)\,\sigma_z(i+1),
  \label{eq:tfim}
\end{equation}
where $\sigma_a(i)$ acts on cell $i$ and each bond is counted once, so that an antiparallel pair costs $\Ek$ more than a parallel one~\cite{toth2001role,taucer2015consequences,retallick2020limits}.
Equation~\eqref{eq:tfim} is the transverse-field Ising model (TFIM), with Ising exchange $J=\Ek/2$ and transverse field $\gamma$.

The intercellular Hartree approximation (ICHA)~\cite{tougaw1996dynamic,toth2001role}, a mean-field approximation, replaces the state of the line by a product of single-cell density matrices and so neglects all intercell correlation.
Each cell then evolves under Eq.~\eqref{eq:hamiltonian}, which follows from Eq.~\eqref{eq:tfim} on replacing $\sigma_z(j)$ by $\langle\sigma_z(j)\rangle=-P_j$.
Writing the density matrix of cell $i$ as $\rho_i=\tfrac12(I+\vlam_i\cdot\vec\sigma)$ defines its coherence vector, with components $\lambda_{i,a}=\langle\sigma_a\rangle_i=\Tr(\rho_i\sigma_a)$: the interdot coherence $\lambda_{i,x}$, an interdot current $\lambda_{i,y}$ and $\lambda_{i,z}=-P_i$.
The equation of motion $\mathrm i\hbar\dot\rho_i=[H_i,\rho_i]$ becomes the precession of the coherence vector about $\vec\Gamma_i$ (Appendix~\ref{app:derivations}), in the form of T\'oth and Lent~\cite{toth2001role}:
\begin{equation}
  \hbar\,\frac{\diff\vlam_i}{\diff t}=\vec\Gamma_i\times\vlam_i .
  \label{eq:precession}
\end{equation}
In components,
\begin{align}
  \hbar\dot\lambda_{i,x}&=-\Eeff\,\lambda_{i,y}, \label{eq:eom_x}\\
  \hbar\dot\lambda_{i,y}&=\Eeff\,\lambda_{i,x}-2\gamma P_i, \label{eq:eom_y}\\
  \hbar\dot P_i&=2\gamma\lambda_{i,y}. \label{eq:eom_z}
\end{align}

\paragraph*{Dissipation.}
Following Timler and Lent~\cite{timler2002power}, coupling to the environment enters as relaxation of each cell toward its instantaneous thermal equilibrium, the coherence-vector counterpart of kinetic Ising models~\cite{glauber1963time,suzuki1968dynamics}:
\begin{align}
  \hbar\,\frac{\diff\vlam_i}{\diff t}&=\vec\Gamma_i\times\vlam_i-\frac\hbar\tau\bigl(\vlam_i-\vlam_{i,\mathrm{ss}}\bigr),
  \label{eq:coherence_eom}\\
  \vlam_{i,\mathrm{ss}}&=-\frac{\vec\Gamma_i}{|\vec\Gamma_i|}\tanh\!\left(\frac{|\vec\Gamma_i|}{2\kB T}\right),
  \label{eq:lambda0}
\end{align}
with relaxation time $\tau$ and temperature $T$; $\vlam_{i,\mathrm{ss}}$ is the coherence vector of the thermal state of $H_i$.
For an isolated cell, $\vec\Gamma=(-2\gamma,0,0)$ and $\vlam_{\mathrm{ss}}=(\theta,0,0)$ with
\begin{equation}
  \theta\equiv\tanh(\gamma/\kB T) ;
  \label{eq:theta}
\end{equation}
at $T=0$, $\lambda_x=+1$: the electron occupies the symmetric superposition of the two configurations, and $P=0$.

\begin{figure*}[t]
\centering
\includegraphics[width=\textwidth]{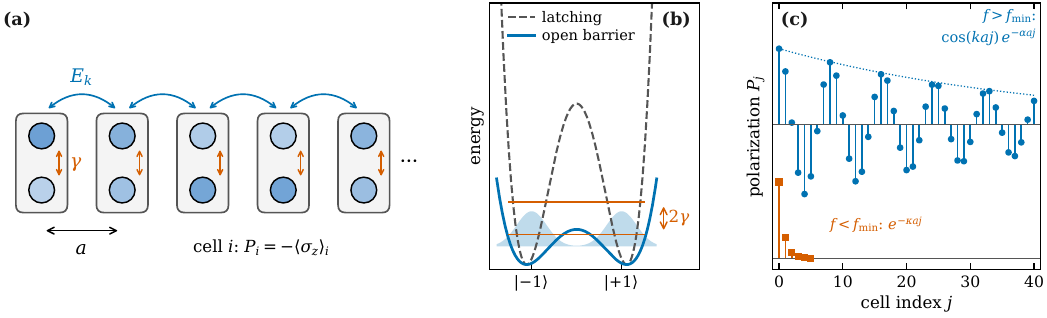}
\caption{Concept.
(a)~A line of two-state cells with pitch $a$, intracell tunnelling $\gamma$ and intercell kink coupling $\Ek$. The shading of the dots shows a small polarization pattern, $P_i=-\langle\sigma_z\rangle_i$, on top of the delocalized, unpolarized state.
(b)~Intracell double well over the two configurations $|{-1}\rangle$ and $|{+1}\rangle$: a high barrier (grey dashed) localizes the electron, while with a low, open barrier (blue) the ground state is delocalized (shaded density) and the two lowest levels are split by $2\gamma$. Whether a line latches is a collective property: for the clean, uniform nearest-neighbour chain at zero temperature, the exact chain latches for $\gamma<\Ek/2$ and is paramagnetic for $\gamma>\Ek/2$.
(c)~Small-amplitude response of the line, $P_j$ versus cell index $j$, for $\gamma=2\Ek$, $\Ek=\SI{1}{meV}$ (Secs.~\ref{sec:ac} and~\ref{sec:damping}). Above the cutoff ($f>\fmin$, blue) a damped travelling wave $\cos(kaj)\,e^{-\alpha aj}$ with $ka=0.245\pi$ and attenuation $\alpha a=0.0236$ per cell for $\tau=40\,\hbar/\Ek$ (dotted: envelope); below it ($f<\fmin$, orange, \SI{50}{GHz}, drawn on a lower baseline) an evanescent response $e^{-\kappa aj}$ with $\kappa a=1.314$.}
\label{fig:schematic}
\end{figure*}

\paragraph*{The open-barrier regime.}
The ratio $\gamma/\Ek$ decides how a line responds.
In the ICHA a uniform line has a self-consistently polarized (latched) ground state only for $\gamma<\Ek$~\cite{toth2000thesis} (Appendix~\ref{app:derivations}); digital QCA works deep in this regime, with a high interdot barrier and $\gamma\ll\Ek$.
We consider instead the open-barrier regime, $\gamma\gtrsim\Ek$, in which the barrier is low and each cell is delocalized: $\vlam=(1,0,0)$ in the ground state and $(\theta,0,0)$ in thermal equilibrium [Fig.~\ref{fig:schematic}(b)].
There the response of Eq.~\eqref{eq:response} is soft: for $\gamma\ge\Ek$ the link gain $\Ek/2\gamma$ is at most $1/2$, the unpolarized line is stable, and a small polarization imposed on one cell neither saturates nor latches but passes linearly to its neighbours.
A line in this regime is an analog medium.
The exact chain is paramagnetic for all $\gamma>\Ek/2$ (Sec.~\ref{sec:exact}), so the regime extends further than the ICHA suggests.

\paragraph*{Small-signal linearization.}
A small signal leaves every cell close to its unpolarized equilibrium: the polarization $P_i$ and the interdot current $\lambda_{i,y}$ stay small, of order $\delta\ll1$.
The field of the neighbours, $\Eeff$, is then small as well, so by Eq.~\eqref{eq:eom_x} the interdot coherence changes only at second order, $\dot\lambda_{i,x}=\mathcal O(\delta^2)$, and keeps its equilibrium value $\theta$ (1 at $T=0$).
With $\lambda_{i,x}$ fixed, Eqs.~\eqref{eq:eom_y} and~\eqref{eq:eom_z} are linear in $P_i$ and $\lambda_{i,y}$, and Sec.~\ref{sec:ac} solves them.

\section{Small-signal AC theory in the ICHA}
\label{sec:ac}

This section solves the linearized ICHA equations for a uniform line; numerical examples use two reference parameter sets, or envelopes (Env~I and II in the figures): \EnvI\ ($\gamma=\SI{2}{meV}$, $\Ek=\SI{1}{meV}$, $a=\SI{9}{nm}$), a cryogenic sub-terahertz case, and \EnvII\ ($\gamma=\SI{44}{meV}$, $\Ek=\SI{18}{meV}$, $a=\SI{4}{nm}$), a high-frequency case.
Both are illustrative parameter sets. \EnvII\ lies near the lower corner of the prospective open-barrier molecular cells of Fig.~\ref{fig:tmax}; \EnvI\ would need nanometre-scale semiconductor cells with an intracell tunnelling energy two to three orders of magnitude larger than the cited gate-defined-double-dot range. Section~\ref{sec:discussion} discusses both kinds of platform.

\subsection{Discrete wave equation and lattice dispersion}
At $T=0$ the small-signal linearization of Sec.~\ref{sec:model} gives $\lambda_{i,x}=1$, and Eq.~\eqref{eq:eom_y} becomes $\hbar\dot\lambda_{i,y}=\Ek(P_{i-1}+P_{i+1})-2\gamma P_i$.
Differentiating Eq.~\eqref{eq:eom_z} and substituting gives the discrete wave equation
\begin{equation}
  \hbar^2\ddot P_i+4\gamma^2P_i-2\gamma\Ek\,(P_{i-1}+P_{i+1})=0 .
  \label{eq:discrete_wave}
\end{equation}
A travelling wave $P_j\propto e^{\mathrm{i}(kja-\omega t)}$ obeys the lattice dispersion relation
\begin{equation}
  \hbar^2\omega^2=4\gamma\bigl(\gamma-\Ek\cos ka\bigr).
  \label{eq:dispersion}
\end{equation}
For $\gamma>\Ek$ the band runs from $\wmin=(2/\hbar)\sqrt{\gamma(\gamma-\Ek)}$ at $k=0$ to $\omega_{\max}=(2/\hbar)\sqrt{\gamma(\gamma+\Ek)}$ at the zone edge, so it is narrow:
\begin{equation}
  \frac{\omega_{\max}}{\wmin}=\sqrt{\frac{\gamma+\Ek}{\gamma-\Ek}} ,
  \label{eq:band_ratio}
\end{equation}
which is $\sqrt3=1.73$ for $\gamma=2\Ek$.
For $\gamma<\Ek$, $\omega^2<0$ at long wavelengths (Sec.~\ref{sec:continuum}).
Equation~\eqref{eq:dispersion} coincides with the random-phase approximation (RPA), or linear spin-wave dispersion, of Eq.~\eqref{eq:tfim} about the $x$-polarized state~\cite{degennes1963collective,brout1966tunnelling,stinchcombe1973ising}; Section~\ref{sec:exact} compares it and its regime boundaries with the exact ones.

\subsection{Cutoff and evanescent band}
The lower band edge in ordinary frequency is
\begin{equation}
  \fmin=\frac{\wmin}{2\pi}=\frac{\sqrt{\gamma(\gamma-\Ek)}}{\pi\hbar} ,
  \label{eq:fmin}
\end{equation}
in the sub-terahertz to terahertz range for meV energies (\SI{684}{GHz} for \EnvI).
As a reference carrier we take $f_c\approx1.13\,\fmin$, which in \EnvI\ is $ka=0.245\pi$, or \SI{774}{GHz}.
Below the cutoff the wavenumber is imaginary, $k=\mathrm{i}\kappa$, and Eq.~\eqref{eq:dispersion} gives
\begin{equation}
  \cosh\kappa a=\frac{\gamma-\hbar^2\omega^2/4\gamma}{\Ek} .
  \label{eq:kappa}
\end{equation}
For \EnvI\ a \SI{50}{GHz} drive therefore decays as $e^{-1.31}\approx0.27$ per cell, and in the exact chain as $e^{-1.38}$ [Fig.~\ref{fig:damping}(b)].
The open-barrier line is thus intrinsically a high-pass medium whose band is set by $\gamma$ and $\Ek$.

\subsection{Continuum limit}
\label{sec:continuum}
Expanding $P_{i\pm1}$ to second order in $a$ turns Eq.~\eqref{eq:discrete_wave} into
\begin{equation}
  \frac{\partial^2P}{\partial t^2}-c^2\frac{\partial^2P}{\partial x^2}+\wmin^2P=0,
  \qquad
  c\equiv\frac a\hbar\sqrt{2\gamma\Ek},
  \label{eq:kg}
\end{equation}
a Klein--Gordon equation for a real scalar field, equivalent to the small-$ka$ expansion $\omega^2\simeq\wmin^2+c^2k^2$ of Eq.~\eqref{eq:dispersion}.
At the cutoff the whole line oscillates in phase, as at the lower band edge of the latched line~\cite{toth2000thesis}, so the Coulomb energy between neighbours is minimal and the excitation energy $\hbar\wmin$ is dominated by tunnelling.
The velocity $c$ is $a$ times the geometric mean of the tunnelling frequency $2\gamma/\hbar$ and the Coulomb frequency $\Ek/\hbar$; for \EnvI, $c=\SI{27.3}{km/s}$.
By analogy with the relativistic case, a packet has the effective rest mass
\begin{equation}
  \meff\equiv\frac{\hbar\wmin}{c^2}=\frac{\hbar^2\sqrt{\gamma(\gamma-\Ek)}}{\gamma\Ek a^2},
  \label{eq:meff}
\end{equation}
which is $0.67\,m_e$ for \EnvI.
Because the band is narrow [Eq.~\eqref{eq:band_ratio}], the relativistic regime $\omega\gg\wmin$ occurs inside the band only close to the critical point.

For $\gamma<\Ek$ the ICHA cutoff is imaginary ($\wmin^2<0$, a tachyonic mass), and for $|ka|<\arccos(\gamma/\Ek)$ perturbations of the unpolarized state grow as $e^{\Omega_It}$, with
\begin{equation}
  \Omega_I(k)=\frac2\hbar\sqrt{\gamma\bigl(\Ek\cos ka-\gamma\bigr)} .
  \label{eq:growth}
\end{equation}
In the ICHA this instability is the spinodal latching of digital QCA: fluctuations grow until the nonlinear dynamics pins the line into saturated domains [Fig.~\ref{fig:wavefront}(d) in Appendix~\ref{app:ed}].
In the exact chain the boundary lies at $\gamma=\Ek/2$ (Sec.~\ref{sec:exact}).

\subsection{Phase and group velocity}
\begin{figure*}[t]
\centering
\includegraphics[width=\textwidth]{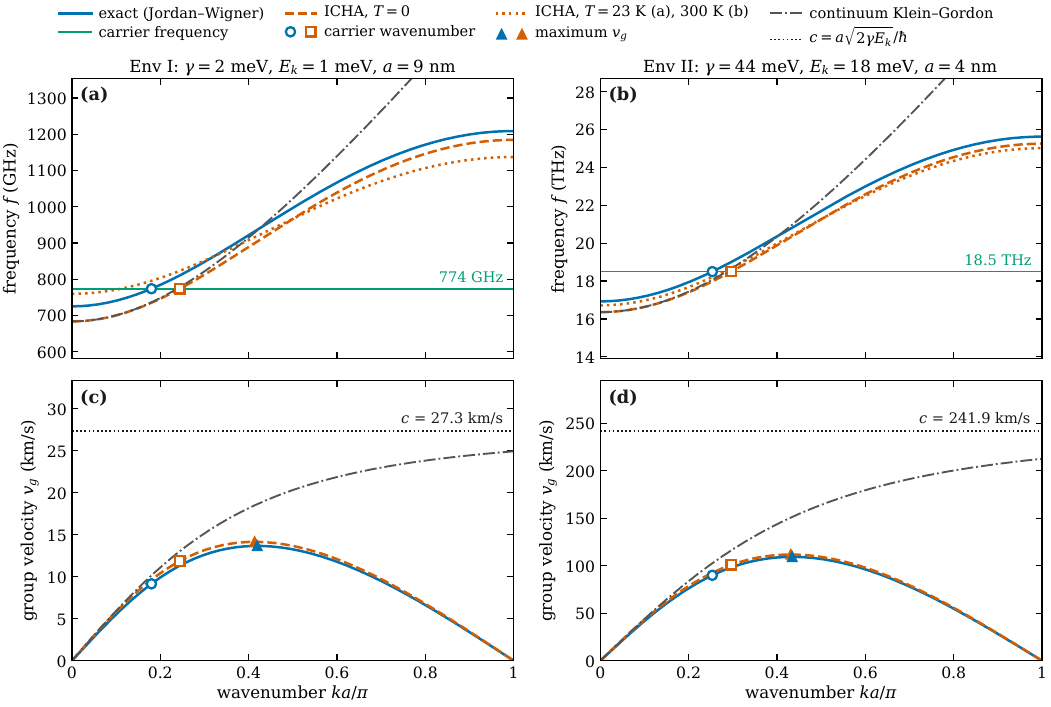}
\caption{Dispersion (a,b) and group velocity (c,d) of the QCA line for \EnvI\ (a,c: $\gamma=\SI{2}{meV}$, $\Ek=\SI{1}{meV}$, $a=\SI{9}{nm}$) and \EnvII\ (b,d: $\gamma=\SI{44}{meV}$, $\Ek=\SI{18}{meV}$, $a=\SI{4}{nm}$).
Curves: exact quasiparticle dispersion $\varepsilon_k$, Eq.~\eqref{eq:tfim_exact} (blue, solid); ICHA at $T=0$, Eq.~\eqref{eq:dispersion} (orange, dashed); ICHA with the thermal coupling $\theta\Ek$ of Sec.~\ref{sec:finiteT} at \SI{23}{K} (a) and \SI{300}{K} (b) (orange, dotted); continuum Klein--Gordon law with the ICHA gap and $c=a\sqrt{2\gamma\Ek}/\hbar$ (grey, dash-dotted). The finite-temperature ICHA curves shift the cutoff through $\theta$, whereas the exact quasiparticle energies do not depend on temperature (Sec.~\ref{sec:finiteT}).
The green line marks the reference carrier (\SI{774}{GHz} and \SI{18.5}{THz}); open markers give its wavenumber (circle exact, square ICHA; $ka=0.180\pi$ and $0.245\pi$ in \EnvI).
In (c,d) the black dotted line is $c$ (\SI{27.3}{km/s} and \SI{241.9}{km/s}), filled triangles mark the maximum $\vg$ (exact $a\Ek/\hbar=13.7$ and \SI{109}{km/s}; ICHA 14.2 and \SI{112}{km/s}), and open markers give $\vg$ at the carrier.}
\label{fig:dispersion}
\end{figure*}

From Eq.~\eqref{eq:dispersion},
\begin{equation}
  \vp=\frac{\omega}{k},
  \qquad
  \vg=\frac{\partial\omega}{\partial k}=\frac{2a\gamma\Ek\sin ka}{\hbar^2\omega(k)} ,
  \label{eq:velocities}
\end{equation}
and on the lattice the two satisfy exactly
\begin{equation}
  \vp\vg=c^2\,\frac{\sin ka}{ka},
  \label{eq:velocity_product}
\end{equation}
which reduces to the relativistic relation $\vp\vg=c^2$ only for $ka\ll1$.
The phase velocity diverges at the cutoff, while the group velocity vanishes there and at the zone edge.
In the ICHA the maximum group velocity depends on $\gamma$; for $\gamma=2\Ek$ it is
\begin{equation}
  \vg^{\max}=1.035\,\frac{a\Ek}{\hbar}\quad(\text{ICHA},\ \gamma=2\Ek),
  \label{eq:vgmax}
\end{equation}
whereas the exact value is $a\Ek/\hbar$ for every $\gamma>\Ek/2$ (Sec.~\ref{sec:exact}): \SI{13.7}{km/s} for \EnvI, about half of $c$.
Figure~\ref{fig:dispersion} compares the ICHA and exact dispersions and group velocities.

\section{Exact benchmark and validity map of the ICHA}
\label{sec:exact}

The chain of Eq.~\eqref{eq:tfim} is exactly solvable, so the mean-field theory of Sec.~\ref{sec:ac} can be tested rather than assumed.
The test gives three results: the exact squared dispersion is the ICHA one shifted by $\Ek^2$, a correction of at most \SI{6}{\percent} in frequency for $\gamma\ge2\Ek$; the wave is a quasiparticle of the interacting chain; and waves propagate down to $\gamma=\Ek/2$, half the tunnelling energy at which the ICHA predicts latching.
Table~\ref{tab:validity} collects them as a validity map.

\begin{table}[t]
\caption{Validity map of the ICHA for small signals on a clean, uniform, zero-temperature line with nearest-neighbour coupling. The frequency bounds and open-band group-velocity bounds for $\gamma\ge2\Ek$ are quantified in Appendix~\ref{app:derivations}; endpoint relative-velocity errors are understood as limits.}
\label{tab:validity}
\begin{ruledtabular}
\begin{tabular}{l>{\raggedright\arraybackslash}p{2.2cm}>{\raggedright\arraybackslash}p{3.8cm}}
Tunnelling & Exact chain & ICHA \\
\hline
$\gamma\ge2\Ek$ & waves; a single quasiparticle & quantitative dispersion: frequency and open-band group velocity within about \SI{6}{\percent} \\[3pt]
$\Ek<\gamma\lesssim2\Ek$ & waves & qualitatively correct; the cutoff lies increasingly below the exact one as $\gamma\to\Ek$ and is taken from Eq.~\eqref{eq:tfim_exact} \\[3pt]
$\Ek/2<\gamma<\Ek$ & waves, with a low, tunable cutoff & fails qualitatively: predicts latching \\[3pt]
$\gamma<\Ek/2$ & ordered & ordered \\
\end{tabular}
\end{ruledtabular}
\end{table}

\subsection{Jordan--Wigner solution}
A Jordan--Wigner transformation maps Eq.~\eqref{eq:tfim} onto free fermions $c_k$~\cite{pfeuty1970one}, $H=\sum_k\varepsilon_k(c^\dagger_kc_k-\tfrac12)$, with
\begin{align}
  \varepsilon_k&=2\sqrt{\gamma^2+\frac{\Ek^2}{4}-\gamma\Ek\cos ka}\,,\nonumber\\
  \varepsilon_k^2&=\hbar^2\omega_k^2+\Ek^2 ,
  \label{eq:tfim_exact}
\end{align}
where $\hbar\omega_k$ is the ICHA/RPA dispersion, Eq.~\eqref{eq:dispersion}: the exact squared dispersion is the ICHA one shifted up by $\Ek^2$.
Equation~\eqref{eq:dispersion} therefore holds for the exact chain with $\hbar^2\omega^2$ replaced by $\hbar^2\omega^2-\Ek^2$, and the evanescent decay and the velocities of Sec.~\ref{sec:ac} follow from it in the same way.
The gap is $\Delta=2|\gamma-\Ek/2|$, the band top is $2(\gamma+\Ek/2)$, and the critical point is $\gamma_c=\Ek/2$, half the ICHA value $\Ek$.
In the ordered phase the quasiparticles are domain walls~\cite{sachdev2011quantum}, the quantum counterparts of the kink solitons of Refs.~\cite{nakagawa2001kink,nakagawa2001junction} and of the tight-binding kink pseudo-particles of Tougaw and Lent~\cite{tougaw1996dynamic}.

\subsection{Comparison with the ICHA theory}
Far from the critical point the two theories agree.
They have the same Klein--Gordon velocity $c$, and their gaps converge as the barrier opens (Appendix~\ref{app:derivations}).
For $\gamma=2\Ek$ the exact cutoff lies \SI{6}{\percent} above the ICHA value (\SI{725}{GHz} against \SI{684}{GHz} for \EnvI), and the band top moves less (Fig.~\ref{fig:dispersion}, Appendix~\ref{app:derivations}).

\begin{figure*}[t]
\centering
\includegraphics[width=\textwidth]{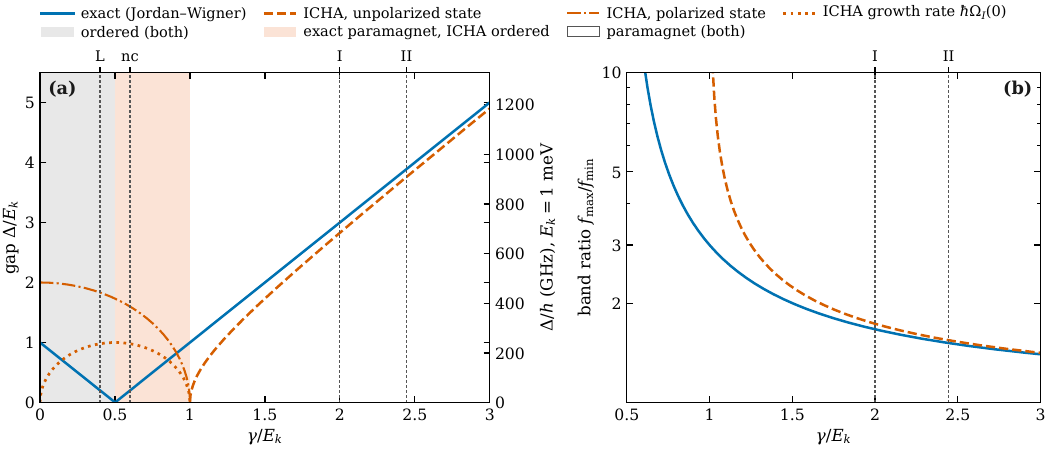}
\caption{Zero-temperature gap phase diagram of the clean, uniform nearest-neighbour chain.
(a)~Excitation gap $\Delta/\Ek$ versus $\gamma/\Ek$; the right axis gives $\Delta/h$ for $\Ek=\SI{1}{meV}$. Curves: exact gap, $\Delta=2|\gamma-\Ek/2|$ (blue, solid); ICHA gap about the unpolarized state, $2\sqrt{\gamma(\gamma-\Ek)}$ (dashed), and about the polarized ICHA state, $2\sqrt{\Ek^2-\gamma^2}$~\cite{toth2000thesis} (dash-dotted); ICHA growth rate $\hbar\Omega_I(0)$ of the unpolarized state, Eq.~\eqref{eq:growth} (dotted).
Shading: grey, ordered in both theories; orange, the near-critical window $\Ek/2<\gamma<\Ek$, where the exact chain is a paramagnet but the ICHA latches; white, paramagnetic in both.
Top ticks: L, $\gamma=0.4\,\Ek$ (a latching example); nc, $\gamma=0.6\,\Ek$ (near-critical); I and II, the two reference parameter sets.
(b)~Ratio of band top to band bottom: exact, $(\gamma+\Ek/2)/(\gamma-\Ek/2)$, and ICHA, $\sqrt{(\gamma+\Ek)/(\gamma-\Ek)}$. The regime $\omega\gg\wmin$ exists inside the band only close to the critical point.}
\label{fig:phase}
\end{figure*}

The exact chain differs from the ICHA in two ways that matter for signals.
First, its maximum group velocity is exactly $a\Ek/\hbar$ for every $\gamma>\Ek/2$, whereas the ICHA maximum depends on $\gamma$ [Eq.~\eqref{eq:vgmax}]; this is the known maximal quasiparticle velocity of the transverse-field Ising chain~\cite{calabrese2011quench,calabrese2012quench,mahoney2024lieb,kaneko2023dynamics} (Appendix~\ref{app:derivations}).
Second, the two theories place the critical point differently (Fig.~\ref{fig:phase}).
The exact gap closes linearly, $\Delta\propto|\gamma-\gamma_c|$, as in the two-dimensional Ising universality class~\cite{suzuki1976relationship,sachdev2011quantum}; the ICHA gap closes as a square root, and at $\gamma=\Ek$ rather than $\Ek/2$.
In the near-critical window $\Ek/2<\gamma<\Ek$ the ICHA therefore predicts an imaginary rest mass and latching [Eq.~\eqref{eq:growth}], whereas the exact chain is a gapped paramagnet that carries waves with a cutoff $(2\gamma-\Ek)/h$, which can be tuned arbitrarily low.

\subsection{The propagating mode in the full quantum model}
Exact diagonalization (ED) of periodic chains with $N=12$--20 cells confirms that the wave is an excitation of the interacting chain (Fig.~\ref{fig:szz}, Appendix~\ref{app:ed}).
For $\gamma=2\Ek$ a single sharp quasiparticle peak on $\varepsilon_k$ carries more than \SI{99.99}{\percent} of the spectral weight of the zero-temperature dynamical structure factor $S^{zz}(k,\omega)$ at every $k$, as expected far from the critical point~\cite{hamer2006critical}.
At $\gamma=0.6\,\Ek$, where the ICHA has no stable unpolarized state, the quasiparticle still carries \SI{98}{\percent} of the weight integrated over $k$.

Time evolution gives the same picture (Fig.~\ref{fig:wavefront}).
After a small local tilt of one cell the polarization develops a ballistic front with asymptotic velocity $\vg^{\max}=a\Ek/\hbar=c\sqrt{\Ek/2\gamma}$, i.e.\ $c/2$ for $\gamma=2\Ek$ (front-speed fits in Appendix~\ref{app:ed}).
At $\gamma=0.6\,\Ek$ the ICHA chain latches into saturated domains, while the exact chain transmits a small-amplitude wave.
The asymptotic ballistic polarization-front velocity is therefore $\vg^{\max}$; threshold-defined arrival times and finite-distance corrections are specified in Appendix~\ref{app:ed}.

\subsection{Validity map}
Table~\ref{tab:validity} summarizes where the ICHA theory can be used.
Taucer et al.\ concluded that the ICHA may be useful at low tunnelling rates, where cells latch~\cite{taucer2015consequences}; for small signals about the unpolarized state it becomes accurate in the opposite limit, because the correlation length falls below one cell.
Its failure in the window $\Ek/2<\gamma<\Ek$ is consistent with the spurious ICHA latching found by Taucer et al.~\cite{taucer2015consequences}.
Close to $\gamma_c=\Ek/2$ the gap closes, and the response becomes sensitive to temperature, which must satisfy $\kB T\ll\Delta$, and to disorder, since local variations of $\gamma$ and $\Ek$ comparable to $\Delta$ close the gap locally~\cite{fisher1995random,mahoney2026lieb}.

\section{Damping and finite temperature}
\label{sec:damping}

This section adds the relaxation term of Eq.~\eqref{eq:coherence_eom} and temperature to the linear theory.

\subsection{Damping toward the instantaneous equilibrium}
To first order in the signal, Eq.~\eqref{eq:lambda0} gives $\vlam_{i,\mathrm{ss}}\approx(\theta,0,-\theta\Eeff/2\gamma)$: the target polarization is $P_{i,\mathrm{ss}}=\theta\Eeff/2\gamma$ and the target current is zero.
Eliminating $\lambda_{i,y}$ (Appendix~\ref{app:derivations}) gives
\begin{multline}
  \hbar^2\ddot P_i+\frac{2\hbar^2}\tau\dot P_i+\Bigl(4\gamma^2+\frac{\hbar^2}{\tau^2}\Bigr)P_i-2\gamma\theta\Eeff\\
  -\frac{\hbar^2\theta}{2\gamma\tau}\dot E_{k,i}^{\mathrm{eff}}-\frac{\hbar^2\theta}{2\gamma\tau^2}\Eeff=0 .
  \label{eq:damped_full}
\end{multline}
Without the relaxation target the last two terms of Eq.~\eqref{eq:damped_full} drop out, and at $T=0$ ($\theta=1$) the continuum limit of the rest is the Telegrapher (damped Klein--Gordon) equation
\begin{equation}
  \frac{\partial^2P}{\partial t^2}+\frac2\tau\frac{\partial P}{\partial t}-c^2\frac{\partial^2P}{\partial x^2}+\omega_0^2P=0,
  \quad
  \omega_0^2\equiv\wmin^2+\tau^{-2},
  \label{eq:telegrapher}
\end{equation}
whose envelope decays at the full rate $1/\tau$, not $1/2\tau$, because the wave is carried jointly by $\lambda_y$ and $P$ in quadrature and both relax.
Keeping the relaxation target couples the damping of each cell to its neighbours, and for a plane wave Eq.~\eqref{eq:damped_full} gives the damping rate
\begin{equation}
  \beta(k)=\frac1\tau\Bigl[1-\frac{\theta\Ek\cos ka}{2\gamma}\Bigr]
  \label{eq:beta_k}
\end{equation}
and the frequency $\hbar^2\omega^2=4\gamma(\gamma-\theta\Ek\cos ka)-(\hbar\theta\Ek\cos ka/2\gamma\tau)^2$.
The spatial attenuation constant and decay length are
\begin{equation}
  \alpha=\frac{\beta}{\vg}=\frac{1-\theta\Ek\cos ka/2\gamma}{\vg\tau},
  \qquad
  \ldecay=\frac1\alpha .
  \label{eq:attenuation}
\end{equation}
This conversion holds for weak damping, $\beta\ll\omega$, away from the band edges (Appendix~\ref{app:derivations}).
For $\gamma=2\Ek$ at the reference carrier the bracket is $0.82$, so $\ldecay$ is $1.22$ times $\vg\tau$.
Simulations of the nonlinear ICHA chain reproduce this attenuation and, below the gap, the evanescence of Eq.~\eqref{eq:kappa}, and ED reproduces the exact evanescence (Fig.~\ref{fig:damping}, Appendix~\ref{app:chain}).

\subsection{Finite temperature}
\label{sec:finiteT}
In the ICHA, temperature enters through the relaxation target: for an isolated open-barrier cell $\vlam_{\mathrm{ss}}=(\theta,0,0)$, so $\theta=\tanh(\gamma/\kB T)$ is the equilibrium interdot coherence $\langle\sigma_x\rangle$ and the equilibrium polarization is zero.
Linearizing about $\lambda_{i,x}=\theta$ renormalizes the coupling to $\theta\Ek$, which raises the ICHA cutoff; for \EnvI\ $\theta$ stays above 0.98 up to \SI{10}{K} (Appendix~\ref{app:derivations}).

The exact chain behaves differently: its quasiparticle energies do not shift with temperature, but thermally excited quasiparticles scatter the signal, at a rate $\tau_{\mathrm{th}}^{-1}\simeq(2\kB T/\pi\hbar)\,e^{-\Delta/\kB T}$ at low temperature~\cite{sachdev1997relaxational,essler2008finite}.
The dilute thermal quasiparticles collide with an excitation and randomize its phase, so the $\sigma_z$ autocorrelation function decays at the rate $\tau_{\mathrm{th}}^{-1}$. A small-signal wave packet is a superposition of the single-quasiparticle excitations created by $\sigma_z$, and we identify $\tau_{\mathrm{th}}$ with the time over which a launched carrier keeps a definite phase relative to its drive.
In the leading stationary low-temperature estimate, $\tau_{\mathrm{th}}$ for \EnvI\ falls from \SI{18}{ns} at \SI{4}{K} to \SI{39}{ps} at \SI{10}{K} and exceeds tens of picoseconds only for $T\lesssim\SI{10}{K}$; the finite-band moving-carrier correction is treated below.

\subsection{Temperature ceiling from thermal dephasing}
\label{sec:tmax}
\begin{figure}[t]
\centering
\includegraphics[width=\columnwidth]{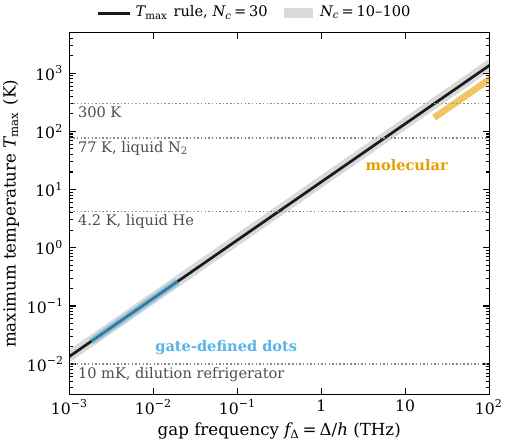}
\caption{Leading stationary low-temperature estimate of the thermal-quasiparticle dephasing ceiling in the nearest-neighbour transverse-field Ising chain, Eq.~\eqref{eq:tmax}, versus the gap frequency $f_\Delta\equiv\Delta/h$ of the exact chain, with the carrier taken at $f_\Delta$. Black line, $N_c=30$ carrier periods (\SI{13.6}{K} per THz); grey band, $N_c=10$--100 (17.8 to \SI{10.7}{K} per THz). Dotted lines: \SI{300}{K}; \SI{77}{K} (liquid nitrogen); \SI{4.2}{K} (liquid helium); \SI{10}{mK} (dilution refrigerator).
Ribbons, offset from the line, span the gaps $\Delta/h$ at two corners of illustrative parameter boxes (Sec.~\ref{sec:discussion}): gate-defined double dots, 1.8--\SI{19}{GHz}, at $(\gamma,\Ek)=(5,2.5)\,\mu$eV and $(50,20)\,\mu$eV (sky blue, on the line); molecular open-barrier cells, 18--\SI{181}{THz}, at $\gamma=2\Ek=0.05$ and \SI{0.5}{eV} (orange, below the line, cut at \SI{100}{THz}).}
\label{fig:tmax}
\end{figure}

As the criterion for a usable carrier we require that it keep a definite phase for $N_c$ periods before thermal quasiparticles dephase it; $N_c$ is set by the application, not by the line, and the estimate below is the leading stationary one.
With $f_\Delta\equiv\Delta/h$ the gap frequency of the exact chain ($\Delta=2\gamma-\Ek$ for nearest-neighbour coupling; it lies above the ICHA cutoff $\fmin$, by \SI{6}{\percent} at $\gamma=2\Ek$), a carrier near the band bottom, $f\approx f_\Delta$, requires $\tau_{\mathrm{th}}\ge N_c/f_\Delta$.
With the rate above and $x=\Delta/\kB T$, the condition reads $x\,e^{x}\ge4N_c$, which holds for $x\ge W(4N_c)$, where $W$ is the principal branch of the Lambert function, so
\begin{equation}
  \kB T_{\max}\approx\frac{\Delta}{W(4N_c)}=\frac{hf_\Delta}{W(4N_c)} ;
  \label{eq:tmax}
\end{equation}
a carrier at $f_c$ replaces $4N_c$ by $4N_cf_\Delta/f_c$.
For $N_c=30$, $W(120)=3.527$ and
\begin{equation}
  T_{\max}\approx\SI{13.6}{K}\times\frac{f_\Delta}{\SI{1}{THz}}\qquad(N_c=30),
  \label{eq:tmax_rule}
\end{equation}
Within this leading estimate, room temperature corresponds to $f_\Delta\gtrsim\SI{22}{THz}$ and liquid helium to $f_\Delta\gtrsim\SI{0.31}{THz}$ (Fig.~\ref{fig:tmax}).

\emph{Regime.}
The ceiling follows from the semiclassical dephasing rate of the nearest-neighbour transverse-field Ising chain in its paramagnetic phase, derived for $\kB T\ll\Delta$; at the ceiling $\Delta/\kB T=W(4N_c)\approx3.5$, and the rate enters $T_{\max}$ only logarithmically, through $W$.
A moving carrier meets thermal quasiparticles at their relative speed, while the finite band changes the stationary normalization. A consistent full-band Boltzmann estimate includes both factors (Appendix~\ref{app:derivations}). At $N_c=30$ it lowers $T_{\max}$ relative to the stationary asymptotic estimate at the same reference carrier by \SI{3.0}{\percent} for \EnvI\ and \SI{3.3}{\percent} for \EnvII. These are parameter-specific corrections within the kinetic approximation.
Longer-range couplings and two-dimensional layouts change the thermal scattering, and other dephasing channels (Sec.~\ref{sec:discussion}) add to the rate.
The single-cell coherence condition $\theta\ge0.9$ is less restrictive than Eq.~\eqref{eq:tmax} (Appendix~\ref{app:derivations}).

The leading gap-carrier estimate depends only on the gap and on $N_c$; the full-band moving-carrier correction also depends on $\Ek/\Delta$ and $f_c$. The leading estimate places lines of gate-defined double dots, with gigahertz gaps, at tens of millikelvin, and prospective molecular lines with gaps of tens of terahertz at hundreds of kelvin, subject to the model and coherence conditions of Sec.~\ref{sec:discussion}.

\section{Junctions: linear-wave impedance and reflection}
\label{sec:junctions}

This section derives the scattering of small-signal waves at junctions and weak links, exactly within the linearized ICHA lattice.

\subsection{Fan-out junction: lattice scattering in the linearized ICHA}
In a circuit the field on cell $i$ is $\Eeff=\sum_jE_{k,ij}P_j$, summed over all cells coupled to $i$.
The small-signal linearization of Sec.~\ref{sec:model} still gives $\lambda_{i,x}=1$ in every cell, the junction included, and eliminating $\lambda_{i,y}$ gives
\begin{equation}
  \hbar^2\ddot P_i=-4\gamma^2P_i+2\gamma\sum_j E_{k,ij}P_j ,
  \label{eq:graph_wave}
\end{equation}
with $E_{k,ij}\to\theta E_{k,ij}$ at finite temperature.
The dynamical matrix is symmetric, so energy and signal flux are conserved, and inside a uniform arm Eq.~\eqref{eq:graph_wave} reduces to Eq.~\eqref{eq:discrete_wave}, whose only roots at fixed $\omega$ are $\pm k$: junctions produce no evanescent near field inside the band.

Let a junction cell (index 0) join $n+1$ semi-infinite arms: one input and $n$ outputs. In QCA terms this is a fan-out with $n$ outputs ($n=2$ a T, $n=3$ a cross; $n=1$ reduces to a straight wire); in graph terms it is a star.
The bond from the junction to the first cell of arm $\alpha$ has energy $\eta_\alpha\Ek$, and all other bonds have $\Ek$.
In each arm we write $P^\alpha_j=A_\alpha e^{-\mathrm ikja}+B_\alpha e^{\mathrm ikja}$ for $j\ge1$.
The equation of cell $j=1$ is satisfied if the continuation of this form to $j=0$ equals $\eta_\alpha P_0$, so $A_\alpha+B_\alpha=\eta_\alpha P_0$.
Combining this with the junction equation $2\cos(ka)\,P_0=\sum_\alpha\eta_\alpha P^\alpha_1$, in which $P^\alpha_1=\eta_\alpha P_0e^{\mathrm ika}-2\mathrm i\sin ka\,A_\alpha$, gives the junction amplitude
\begin{equation}
  P_0=\frac{2\mathrm i\sin ka\sum_\beta\eta_\beta A_\beta}{\Sigma e^{\mathrm ika}-2\cos ka},\qquad\Sigma=\sum_\alpha\eta_\alpha^2 ,
  \label{eq:p0}
\end{equation}
and, with $B_\alpha=\eta_\alpha P_0-A_\alpha$, the scattering matrix $B=\mathcal SA$ of the linearized ICHA lattice,
\begin{align}
  \mathcal S_{\alpha\beta}&=\eta_\alpha\eta_\beta\,G(k)-\delta_{\alpha\beta},\nonumber\\
  G(k)&=\frac{2}{\Sigma+\mathrm i\,(2-\Sigma)\cot ka},
  \label{eq:smatrix}
\end{align}
where $G$ is the junction function; $\mathcal S$ is unitary.
For the ordinary junction, with all couplings equal to $\Ek$ ($\Sigma=n+1$), the reflection and the transmission into each output are
\begin{align}
  r&=\frac{n-1}{e^{-2\mathrm ika}-n},\qquad
  t=\frac{1-e^{2\mathrm ika}}{1-n\,e^{2\mathrm ika}},\nonumber\\
  |r|^2&=\frac{(n-1)^2}{(n-1)^2+4n\sin^2ka},
  \label{eq:r_uniform}
\end{align}
with $|r|^2+n|t|^2=1$; a straight wire ($n=1$) gives $r=0$.

\emph{Comparison with the transmission-line picture.}
A transmission-line (TL) node whose arms have admittances proportional to $\eta_\alpha^2$ gives $G=2/\Sigma$ and the reflection coefficient $\Gamma_R=(1-n)/(1+n)$ for identical arms: $-1/3$ for a T and $-1/2$ for a cross.
Equation~\eqref{eq:smatrix} is that node loaded by a reactive shunt of normalized susceptance $(\Sigma-2)\cot ka$, which accounts for the extra bonds of the junction cell.
Hence (i) the TL value holds for an ordinary junction only at mid-band, $ka=\pi/2$; (ii) it holds at every $ka$ for a coordination-matched junction with $\Sigma=2$, for example a cross whose four junction bonds are $\Ek/\sqrt2$, the matching found for spin waves on star-shaped spin networks~\cite{yang2006beam}; and (iii) for an ordinary junction the shunt diverges at both band edges, so $r\to-1$ and the junction becomes opaque near the cutoff and near the band top.
The TL result is the mid-band value of a real junction, not its long-wavelength limit (Appendix~\ref{app:junction}).
For a cross, $|r|^2$ is $1/4$ at mid-band and $0.41$ at the reference carrier, and its average over $ka$ is exactly $(n-1)/(n+1)=|\Gamma_R|$, not $\Gamma_R^2$ [Fig.~\ref{fig:junction_rt}(a,b) in Appendix~\ref{app:junction}].
For $\Sigma>2$ the junction also binds one mode below the cutoff and one above the band top; the lower one softens first, so in the ICHA a T or a cross latches at a larger $\gamma$ than a straight wire does (Appendix~\ref{app:junction}).

\subsection{Weak links}
A straight wire in which a single bond is changed to $\eta\Ek$ is the $n=1$ case of Eq.~\eqref{eq:smatrix} with bonds $(1,\eta)$:
\begin{equation}
  |r|^2=\frac{(1-\eta^2)^2}{(1-\eta^2)^2+4\eta^2\sin^2ka},
  \label{eq:defect}
\end{equation}
independent of the sign of $\eta$, with the band average $|1-\eta^2|/(1+\eta^2)$.

\begin{figure*}[tbp]
\centering
\includegraphics[width=\textwidth]{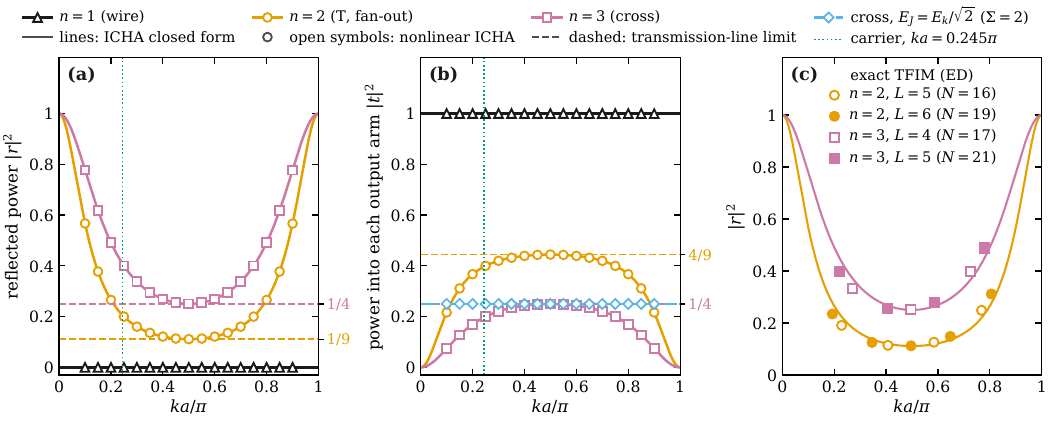}
\caption{Junction scattering on the lattice ($\gamma=2\Ek$) for a straight wire ($n=1$, black, $\blacktriangle$), a T or fan-out ($n=2$, orange, $\circ$) and a cross ($n=3$, purple, $\square$); sky blue, dash-dotted, $\diamond$: a coordination-matched cross with junction bonds $E_J=\Ek/\sqrt2$ ($\Sigma=2$; $E_J$ denotes the junction bond).
Lines are the ICHA closed form, Eq.~\eqref{eq:r_uniform}; open symbols the nonlinear ICHA simulation (17-tone steady state, $|P|\lesssim0.02$); dashed lines the transmission-line (TL) limit, labelled on the right axis. The green dotted line marks the reference carrier, $ka=0.245\pi$.
(a)~Reflected power $|r|^2$ versus $ka/\pi$; TL levels $1/9$ ($n=2$) and $1/4$ ($n=3$).
(b)~Power $|t|^2$ into each output arm; TL levels $4/9$ and $1/4$. The matched cross follows its TL value at every $ka$; it is drawn in (b) only, because in (a) it coincides with the $1/4$ TL line.
(c)~$|r|^2$ from ED of the TFIM on star graphs: $n=2$ with arm length $L=5$ ($N=16$) and $L=6$ ($N=19$); $n=3$ with $L=4$ ($N=17$) and $L=5$ ($N=21$); filled symbols, the largest $L$. Lines: ICHA closed form.}
\label{fig:junction_rt}
\end{figure*}

\emph{Checks.} Nonlinear ICHA simulations of fan-out junctions and weak links reproduce Eqs.~\eqref{eq:smatrix}--\eqref{eq:defect} to $4\times10^{-4}$ in power, and ED of the TFIM on T and cross graphs at $\gamma=2\Ek$ agrees with Eq.~\eqref{eq:r_uniform} to within \SI{16}{\percent} of the reflected power [Fig.~\ref{fig:junction_rt}(c), Appendix~\ref{app:junction}].

\section{Discussion and outlook}
\label{sec:discussion}

\emph{Physical platforms.}
The specified nearest-neighbour two-state model carries waves for $\gamma>\Ek/2$ in the exact chain, while its ICHA dispersion is quantitative for $\gamma\gtrsim2\Ek$ (Sec.~\ref{sec:exact}); propagation also needs a relaxation time of tens of carrier periods. A physical implementation must realize that effective model and its coherence assumptions.
Applying the exact-chain thresholds to a platform requires a controlled reduction of its cells to this nearest-neighbour two-state form: additional cell levels, lattice relaxation and couplings beyond nearest neighbours must be included before that reduction is made, and longer-range couplings also change the junction formulas of Sec.~\ref{sec:junctions}.

\emph{Excitation and readout.}
Driving $\delta P=0.1$ in \EnvI\ requires a bias of \SIrange{0.14}{0.40}{meV} per cell, which calls for local electrodes.
The wave amplitude is an expectation value, while a projective charge measurement returns $\pm1$ with variance $1-P^2$, so resolving a small amplitude $\delta P$ in one cell takes of order $(\delta P)^{-2}$ repetitions or an equivalent ensemble, and weak continuous measurement does not avoid this~\cite{clerk2010noise}.
Radio-frequency charge readout, with carriers of about 0.1--\SI{2}{GHz} and bandwidths of about \SI{100}{MHz}~\cite{wahlgren1999radio,reilly2015engineering}, or \SI{10}{MHz} single-shot~\cite{buehler2005single}, operates far below the sub-terahertz and terahertz bands of lines with meV-scale couplings, so readout there relies on repeated or ensemble measurements.

\emph{Relaxation and temperature.}
Wave operation needs $\tau$ of tens of carrier periods, tens of picoseconds in \EnvI, so the relaxation time must be controlled rather than minimized; stronger electrode coupling shortens it but also screens $\Ek$.
Phonons and the noise of nearby gates, which limit the charge coherence of single double dots~\cite{hayashi2003coherent,petersson2010quantum}, add dephasing that Eq.~\eqref{eq:tmax} does not contain and lower the ceiling further.

\begin{figure*}[tbp]
\centering
\includegraphics[width=\textwidth]{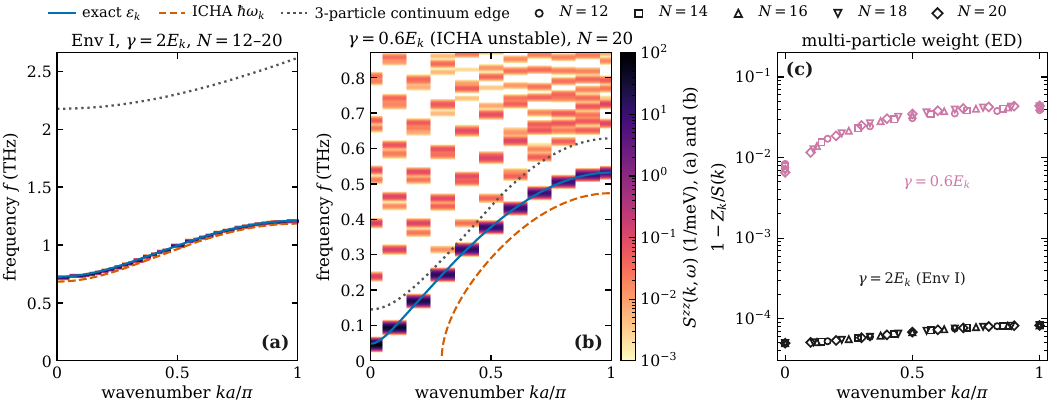}
\caption{Exact dynamical structure factor $S^{zz}(k,\omega)$ of Eq.~\eqref{eq:tfim} at $T=0$, from ED with Lanczos continued fractions (log colour scale, Gaussian broadening $0.02\,\Ek$ for display). Frequencies assume $\Ek=\SI{1}{meV}$.
(a)~$\gamma=2\Ek$ (\EnvI), momenta from $N=12$--20.
(b)~$\gamma=0.6\,\Ek$, $N=20$; the ICHA is unstable here for $|ka|<0.295\pi$, where its curve is absent.
In (a,b) the blue line is $\varepsilon_k$, the orange dashed line the ICHA $\hbar\omega_k$, and the grey dotted line the lower edge of the three-quasiparticle continuum; the colour bar (in meV$^{-1}$) applies to both.
(c)~Multi-particle weight $1-Z_k/S(k)$, with $S(k)=\int S^{zz}(k,\omega)\,\diff\omega$ the static structure factor: the fraction of the spectral weight outside the single-quasiparticle pole, for $N=12$--20 (one marker shape per $N$); black, $\gamma=2\Ek$; purple, $\gamma=0.6\,\Ek$; both exact.}
\label{fig:szz}
\end{figure*}

\emph{Tunable cutoff.}
For $\Ek/2<\gamma<\Ek$ the cutoff is $(2\gamma-\Ek)/h$; at $\gamma=0.6\,\Ek$ with $\Ek=\SI{1}{meV}$ the band runs from \SIrange{48}{532}{GHz}, and the mode remains a well-defined quasiparticle [Sec.~\ref{sec:exact}, Fig.~\ref{fig:szz}(b)].
The leading stationary estimate of Eq.~\eqref{eq:tmax} gives $T\lesssim\SI{0.7}{K}$ in this regime, before moving-carrier and other dephasing corrections, and low disorder is required.

\emph{Gate-defined and molecular cells.}
Gate-defined double dots, whose tunnelling is set by barrier gates and can be tuned in situ~\cite{hayashi2003coherent,petersson2010quantum}, show tunnel couplings $\gamma\approx$ \SIrange{3}{15}{\micro eV} and charge coherence from \SI{0.13}{ns} away from the charge-degeneracy point to 1--\SI{7}{ns} at it~\cite{hayashi2003coherent,petersson2010quantum,shi2013coherent}.
As an illustrative target beyond the cited coupling range, a chain with $\gamma=2\Ek=\SI{20}{\micro eV}$ at a \SI{150}{nm} pitch would carry a 7.3--\SI{12.1}{GHz} band, with a leading stationary temperature estimate of order \SI{100}{mK} (Fig.~\ref{fig:tmax}), provided that coherence and a spread of $2\gamma$ small compared with $2\Ek$ hold along the chain; such chains are a tunable test bed for the theory.
The mixed-valence molecular cells analysed so far are strongly localized by nuclear relaxation, with kink energies of about \SI{0.6}{eV}~\cite{lent2003molecular}, so they sit in the high-barrier regime; a molecular open-barrier line would need an intramolecular tunnelling energy comparable to $\Ek$ and would then operate at tens of terahertz, with the leading temperature estimates shown in Fig.~\ref{fig:tmax}.
Metal-island cells~\cite{orlov1997realization,amlani1999digital} move charge by sequential, incoherent tunnelling, so no coherent splitting forms and the theory does not apply.

\begin{figure*}[tbp]
\centering
\includegraphics[width=\textwidth]{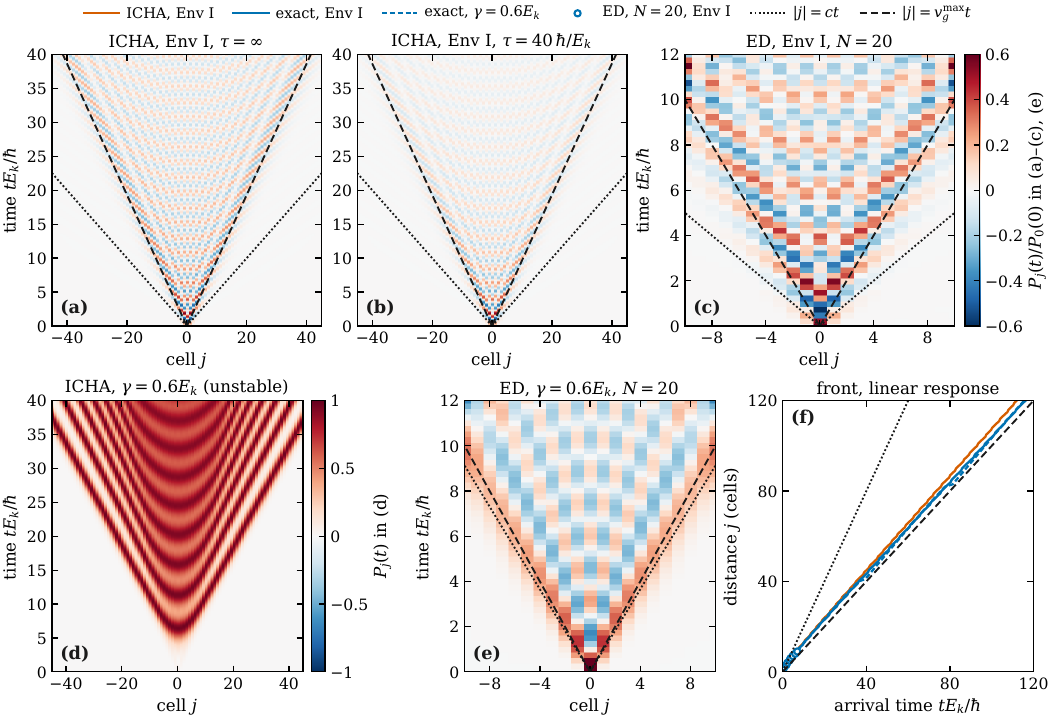}
\caption{Wavefront after a small local polarization tilt (angle $\phi_0=0.05$) of cell $j=0$; time in units of $\hbar/\Ek$.
(a)~Nonlinear ICHA chain, Eq.~\eqref{eq:coherence_eom}, \EnvI\ ($\gamma=2\Ek$), $\tau=\infty$, $N=201$.
(b)~The same with $\tau=40\,\hbar/\Ek$.
(c)~Exact evolution of the quantum chain (ED), \EnvI, $N=20$, periodic.
(d)~ICHA chain at $\gamma=0.6\,\Ek$: the unpolarized state is unstable and the tilt grows into saturated domains (colour shows $P_j$ on its own $\pm1$ scale).
(e)~Exact evolution at $\gamma=0.6\,\Ek$, $N=20$: a small-amplitude wave propagates.
In (a)--(c) and (e) colour shows $P_j(t)/P_0(0)$ on a common scale; dotted lines mark $|j|=ct$ ($c=2\,a\Ek/\hbar$ in \EnvI, $\sqrt{1.2}\,a\Ek/\hbar$ at $0.6\,\Ek$), and dashed lines mark $|j|=\vg^{\max}t$ (ICHA $1.035\,a\Ek/\hbar$, exact $1.000\,a\Ek/\hbar$).
(f)~Arrival time of the front, taken where $|P_j|$ first reaches \SI{3}{\percent} of the initial tilt, versus distance, from infinite-chain linear response: ICHA \EnvI\ (orange), exact \EnvI\ (blue), exact $\gamma=0.6\,\Ek$ (blue, short dashes), ED $N=20$ \EnvI\ (open circles, $j\le9$, near the origin), and the lines $|j|=ct$ (dotted) and $|j|=a\Ek t/\hbar$ (dashed).}
\label{fig:wavefront}
\end{figure*}
\begin{figure*}[tbp]
\centering
\includegraphics[width=\textwidth]{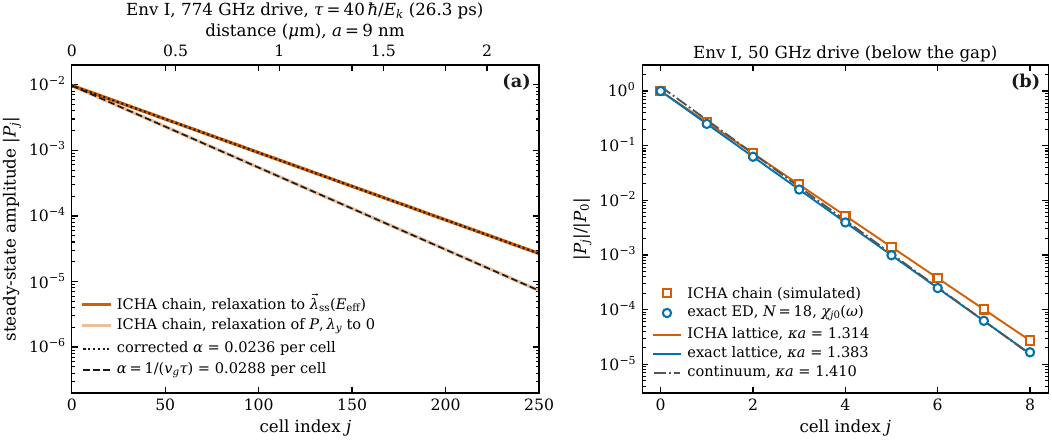}
\caption{Damping and evanescence for \EnvI\ ($\gamma=2\Ek$, $\Ek=\SI{1}{meV}$, $a=\SI{9}{nm}$).
(a)~Steady-state amplitude $|P_j|$ along a 400-cell ICHA chain driven at \SI{774}{GHz} on cell 0 with $\tau=40\,\hbar/\Ek$ (\SI{26.3}{ps}); the top axis gives the distance in \si{\micro m}. Relaxation toward the instantaneous equilibrium $\vlam_{i,\mathrm{ss}}(\Eeff)$ (orange) gives $\alpha=0.0236$ per cell, the corrected value of Eq.~\eqref{eq:attenuation} (black dotted fit); relaxation of $P$ and $\lambda_y$ toward zero (light orange) gives the uncorrected $1/(\vg\tau)=0.0288$ per cell (black dashed fit).
(b)~Below-gap response at \SI{50}{GHz}, $|P_j|/|P_0|$: simulated ICHA chain (squares) and exact susceptibility $\chi_{j0}(\omega)$ from ED with $N=18$ (circles). The lattice decay constants are $\kappa a=1.314$ (ICHA, orange) and $1.383$ (exact, blue); the continuum formula gives $1.410$ (grey, dash-dotted).}
\label{fig:damping}
\end{figure*}

\section{Conclusion}
\label{sec:conclusion}

In the open-barrier regime a QCA line is a linear, dispersive medium for polarization waves.
Its ICHA lattice dispersion has a native cutoff and an evanescent band, reduces to a Klein--Gordon equation at small $ka$, and is damped at the rate $\beta(k)$ of relaxation toward the instantaneous equilibrium.
The exact solution confirms the mode as a sharp quasiparticle, with ICHA frequencies within \SI{6}{\percent} of the exact ones for $\gamma\ge2\Ek$, identifies the asymptotic ballistic-front velocity with the known maximal quasiparticle velocity $a\Ek/\hbar$, and extends the wave regime to $\gamma>\Ek/2$, where the ICHA predicts latching.
The leading stationary low-temperature dephasing estimate gives $\kB T_{\max}\approx\Delta/W(4N_c)$, about \SI{13.6}{K} per terahertz of gap for $N_c=30$; the finite-band moving-carrier correction depends additionally on the bandwidth and carrier.
Within the linearized ICHA, junctions scatter the waves through a lattice matrix that reduces to the transmission-line value only at mid-band.
These closed-form expressions describe the specified nearest-neighbour model and relaxation closure, with the exact-chain comparison setting the observable-specific bounds stated above.
Gate-defined double-dot chains offer a tunable gigahertz test bed.

\section*{Data and code availability}
The code that reproduces the figures and numbers of this article is available from the author on reasonable request.

\begin{acknowledgments}
\emph{Use of AI tools.} AI tools including Claude (Anthropic, including Claude Opus 5.5), GLM-5.3 (Zhipu AI), Grok (xAI) and Codex (OpenAI) assisted in reviewing the derivations, formulating parts of the analysis and the numerical examples, writing the simulation and verification code and editing the manuscript. The author directed the work, verified the AI-assisted content and takes full responsibility for it.
\end{acknowledgments}

\appendix

\section{Derivations}
\label{app:derivations}

\emph{Pauli algebra and the precession.}
With the matrices of Eq.~\eqref{eq:pauli}, $\sigma_x\sigma_y=\mathrm i\,\mathrm{diag}(-1,1)=\mathrm i\sigma_z$, so $[\sigma_a,\sigma_b]=2\mathrm i\,\epsilon_{abc}\sigma_c$, with $\epsilon_{abc}$ the Levi-Civita symbol: changing the sign of both $\sigma_y$ and $\sigma_z$ is a rotation by $\pi$ about $x$ and preserves the algebra.
For $H_i=\tfrac12\vec\Gamma_i\cdot\vec\sigma$ and $\rho_i=\tfrac12(I+\vlam_i\cdot\vec\sigma)$,
\begin{equation}
  [H_i,\rho_i]=\tfrac14\Gamma_a\lambda_b[\sigma_a,\sigma_b]=\tfrac{\mathrm i}2(\vec\Gamma_i\times\vlam_i)\cdot\vec\sigma ,
\end{equation}
and $\mathrm i\hbar\dot\rho_i=\tfrac{\mathrm i\hbar}2\dot\vlam_i\cdot\vec\sigma$ gives $\hbar\dot\vlam_i=\vec\Gamma_i\times\vlam_i$, Eq.~\eqref{eq:precession}.
The components follow from $\vec\Gamma\times\vlam=(-\Eeff\lambda_y,\;\Eeff\lambda_x-2\gamma P,\;-2\gamma\lambda_y)$ with $\lambda_z=-P$, which gives Eqs.~\eqref{eq:eom_x}--\eqref{eq:eom_z}; eliminating $\lambda_y$ gives $\hbar^2\ddot P_i=2\gamma\Eeff\lambda_{i,x}-4\gamma^2P_i$.
Two conventions differ in related work: Timler and Lent~\cite{timler2002power} write Eq.~\eqref{eq:precession} with $1/\hbar$ included in $\vec\Gamma$, and the earlier work of Retallick and the present author~\cite{retallick2020limits} writes the tunnelling term of Eq.~\eqref{eq:tfim} as $-\tfrac12\gamma\sigma_x$, so its $\gamma$ is twice ours.

\emph{Cell-to-cell response.}
The ground state of $H_i=\tfrac12\vec\Gamma_i\cdot\vec\sigma$ has $\vlam=-\vec\Gamma_i/|\vec\Gamma_i|$, so $P=-\lambda_z=\Eeff/\sqrt{4\gamma^2+(\Eeff)^2}$.
One driver gives $\Eeff=\Ek P_{\mathrm{in}}$ and hence Eq.~\eqref{eq:response}.
In a uniform line both neighbours carry $P$, so $\Eeff=2\Ek P$ and $P=(\Ek P/\gamma)/\sqrt{1+(\Ek P/\gamma)^2}$, which has a solution $P\neq0$ only for $\Ek/\gamma>1$: the ICHA line latches for $\gamma<\Ek$~\cite{toth2000thesis}.
At finite temperature the right-hand side is multiplied by $\tanh(|\vec\Gamma|/2\kB T)$, and the threshold becomes $\theta\Ek>\gamma$.

\emph{Linearization with relaxation.}
To first order, Eq.~\eqref{eq:lambda0} gives $\vlam_{i,\mathrm{ss}}\approx(\theta,0,-\theta\Eeff/2\gamma)$, because $|\vec\Gamma_i|=2\gamma+\mathcal O(\delta^2)$ and the field direction tilts by $\Eeff/2\gamma$.
With $\lambda_{i,x}=\theta$, the $y$ and $z$ components of Eq.~\eqref{eq:coherence_eom} become
\begin{align}
  \hbar\dot P_i&=2\gamma\lambda_{i,y}-\frac\hbar\tau\Bigl(P_i-\frac{\theta\Eeff}{2\gamma}\Bigr), \label{eq:damped_p}\\
  \hbar\dot\lambda_{i,y}&=\theta\Eeff-2\gamma P_i-\frac\hbar\tau\lambda_{i,y}. \label{eq:damped_y}
\end{align}
Differentiating Eq.~\eqref{eq:damped_p} and eliminating $\lambda_{i,y}$ and $\dot\lambda_{i,y}$ with Eqs.~\eqref{eq:damped_p} and~\eqref{eq:damped_y} gives Eq.~\eqref{eq:damped_full}.
For $P_j\propto e^{\mathrm{i}kja-\mathrm{i}\omega t-\beta t}$, $\Eeff=2\Ek\cos(ka)\,P_i$.
Writing $s_\omega=-\mathrm{i}\omega-\beta$ and $\mu_k=\theta\Ek\cos ka/\gamma$, Eq.~\eqref{eq:damped_full} becomes
\begin{equation}
  \hbar^2s_\omega^2+\frac{2\hbar^2}{\tau}\Bigl(1-\frac{\mu_k}2\Bigr)s_\omega+4\gamma^2(1-\mu_k)+\frac{\hbar^2}{\tau^2}(1-\mu_k)=0 .
  \label{eq:charpoly}
\end{equation}
In the wave regime, where the discriminant is negative, its roots have real part $-\beta=-\tau^{-1}(1-\mu_k/2)$, which is Eq.~\eqref{eq:beta_k}, and imaginary part given by the frequency quoted below Eq.~\eqref{eq:beta_k}; in the overdamped case both roots are real and their mean is $-\beta$.
Dropping the target terms, i.e.\ setting $\mu_k\to0$ in the coefficient of $s_\omega$ and in the $\hbar^2/\tau^2$ term, recovers the Telegrapher form, Eq.~\eqref{eq:telegrapher}.
At the reference carrier the decay-length factor $1/(1-\theta\Ek\cos ka/2\gamma)$ is 1.22 for $\gamma=2\Ek$ ($\cos ka=0.72$) and 1.14 for $\gamma=2.44\,\Ek$, and for $\omega\tau\approx100$--250 the $\hbar^2/\tau^2$ terms are at most $10^{-4}$ of the inertial term.
Near the band edges, where $\vg\to0$, the conversion $\alpha=\beta/\vg$ of Eq.~\eqref{eq:attenuation} no longer holds, and $\alpha$ is the imaginary part of the complex root $k$ of the damped dispersion at real $\omega$. At finite temperature $\vg$ belongs to the $\theta$-renormalized dispersion.

\emph{Maximum exact group velocity.}
From Eq.~\eqref{eq:tfim_exact}, $\hbar v_g=2\gamma\Ek a\sin ka/\varepsilon_k$, which is Eq.~\eqref{eq:velocities} with the exact frequency in place of the ICHA one.
Maximizing $\sin^2ka/(\gamma^2+\Ek^2/4-\gamma\Ek\cos ka)$ over $u=\cos ka$ gives $\gamma\Ek u^2/2-(\gamma^2+\Ek^2/4)u+\gamma\Ek/2=0$, whose root in $[-1,1]$ for $\gamma>\Ek/2$ is $u=\Ek/2\gamma$.
There the ratio equals $1/\gamma^2$, so $\vg^{\max}=a\Ek/\hbar$ independently of $\gamma$.
This maximal quasiparticle velocity of the transverse-field Ising chain is known from studies of its light cone~\cite{calabrese2011quench,calabrese2012quench}, has been obtained for QCA chains from a Lieb--Robinson correlation function~\cite{mahoney2024lieb}, and is reproduced by linear spin-wave theory only in the strong-field limit~\cite{kaneko2023dynamics}.

\emph{ICHA and exact dispersions.}
The squared dispersions have the same $k^2$ coefficient, $\varepsilon_k^2\simeq\Delta^2+(\hbar ck)^2$ and $\hbar^2\omega_k^2\simeq\hbar^2\wmin^2+(\hbar ck)^2$ with the same $c$, and both dispersions satisfy Eq.~\eqref{eq:velocity_product}; their curvatures at $k=0$ differ because their gaps differ.
For $\gamma\gg\Ek$ the gaps agree to first order, $\Delta-\hbar\wmin=\Ek^2/4\gamma+\mathcal O(\Ek^3/\gamma^2)$.
For $\gamma\ge2\Ek$ the ICHA frequency lies within \SI{5.72}{\percent} of the exact one at every $k$; in the open band $0<|ka|<\pi$ its group-velocity magnitude lies within \SI{6.07}{\percent}. The frequency bound is reached at $\gamma=2\Ek,k=0$, and the relative-velocity bound is approached as $k\to0$; both velocities vanish at the endpoints.
The ICHA peak group velocity, reached near $ka\approx0.41\pi$ for $\gamma=2\Ek$, lies within \SI{3.53}{\percent} of $a\Ek/\hbar$, the bound again reached at $\gamma=2\Ek$.
At $\gamma=2\Ek$ the band top moves by \SI{2}{\percent}, and the ICHA phase velocity exceeds $c$ for $ka<0.74\pi$ and falls to $0.78\,c$ at the zone edge; for $\gamma=2.44\,\Ek$ (\EnvII) the cutoff and the band top shift by \SI{3.5}{\percent} and \SI{1.5}{\percent}.

\emph{Envelope I at finite temperature.}
For \EnvI, $\theta=0.99998$ at \SI{4}{K} ($\fmin=\SI{684}{GHz}$), 0.98 at \SI{10}{K} (\SI{690}{GHz}), 0.91 at \SI{15}{K} (\SI{713}{GHz}) and 0.77 at \SI{23}{K} (\SI{760}{GHz}). The leading stationary asymptotic rate of Sec.~\ref{sec:finiteT} gives $\tau_{\mathrm{th}}\approx\SI{18}{ns}$ at \SI{4}{K}, \SI{39}{ps} at \SI{10}{K} and \SI{8}{ps} at \SI{15}{K}.

\emph{Single-cell coherence.}
For the leading stationary estimate, the condition $\theta\ge0.9$, i.e.\ $\kB T\le\gamma/1.47$, does not compete with Eq.~\eqref{eq:tmax}: because $\Delta\le2\gamma$, $\kB T_{\max}\le2\gamma/W(4N_c)$, which lies below $\gamma/1.47$ for $N_c\ge14$. Corrected moving-carrier roots must be checked under their own conditions.

\emph{Moving-frame dephasing.}
Write $q=ka$, $\varepsilon(q)=\varepsilon_{k=q/a}$ and $v(q)=(a/\hbar)\,\partial_q\varepsilon(q)$. In a full-band Boltzmann collision estimate, a carrier moving at $v_c$ has
\[
 \begin{aligned}
 I_v(T)&=\int_{-\pi}^{\pi}\!\diff q\,e^{-[\varepsilon(q)-\Delta]/\kB T}|v(q)-v_c|,\\
 \Gamma_v&=\frac{e^{-\Delta/\kB T}}{\pi a}I_v(T).
 \end{aligned}
\]
For the paramagnetic nearest-neighbour band, $\varepsilon(\pi)-\Delta=2\Ek$, so changing the integration variable from $q$ to energy gives
\[
 \begin{aligned}
 I_0(T)&=\frac{2a\kB T}{\hbar}F_{\mathrm{band}}(T),\\
 F_{\mathrm{band}}(T)&=1-e^{-2\Ek/\kB T}.
 \end{aligned}
\]
Defining $R_{\mathrm{band}}=I_v/I_0$, the full-band moving rate relative to the asymptotic stationary rate of Sec.~\ref{sec:tmax} is $F_{\mathrm{band}}R_{\mathrm{band}}$, not $R_{\mathrm{band}}$ alone.
Near the band bottom $\varepsilon(q)\simeq\Delta+\hbar^2c^2k^2/2\Delta$, the thermal velocities are Gaussian with variance $v_T^2=c^2\kB T/\Delta$. In this band-bottom approximation the moving/stationary ratio is $R(\zeta)=e^{-\zeta^2/2}+\sqrt{\pi/2}\,\zeta\,\mathrm{erf}(\zeta/\sqrt2)\ge1$, with $\zeta=v_c/v_T$; this ratio does not include the finite-band normalization.
The full-band ceiling is solved self-consistently from
\[
 x e^x\ge4N_c\frac{f_\Delta}{f_c}F_{\mathrm{band}}(T)R_{\mathrm{band}}(T,v_c),
 \qquad x=\frac{\Delta}{\kB T},
\]
using the exact quasiparticle velocity at the physical reference-carrier frequency. For \EnvI\ the retained carrier is defined by the ICHA value $ka=0.245\pi$, giving $f_c=\SI{774.323}{GHz}$; for \EnvII\ $f_c=1.13\,\fmin=\SI{18.4831}{THz}$. At $N_c=10$, 30 and 100 the corrected temperatures are 13.026, 9.714 and \SI{7.549}{K} for \EnvI, and 307.440, 227.113 and \SI{175.718}{K} for \EnvII. Relative to the asymptotic stationary estimate at the \emph{same carrier}, the decreases are 0.9, 3.0 and \SI{3.8}{\percent}, and 0.4, 3.3 and \SI{4.4}{\percent}, respectively.
These are selected-case results of the Boltzmann kinetic approximation, not universal thermal bounds. The gap-carrier line in Fig.~\ref{fig:tmax} uses $f_c=f_\Delta$ and is a different baseline: for $N_c=30$ it gives 9.870 and \SI{230.311}{K} for the two envelopes. Independently written quadratures agree at the quoted precision. At the selected roots $\Delta/\kB T$ ranges from 2.64 to 4.62, so numerical convergence does not by itself control the dilute low-temperature approximation, particularly for $N_c=10$, or establish a device temperature limit.

\section{Exact diagonalization and free-fermion methods}
\label{app:ed}

The simulation code of Appendices~\ref{app:ed}--\ref{app:junction} was written with AI assistance (see Acknowledgments) and checked as described there.
Equation~\eqref{eq:tfim} contains $\sigma_z$ only in pairs, so it does not depend on the sign convention for $\sigma_z$.
The Hamiltonian~\eqref{eq:tfim} is diagonalized matrix-free on periodic chains of $N=12$--20 sites, with spin flips implemented as tensor-axis flips.
The ground state is found by Lanczos iteration with full reorthogonalization, from a random start vector with a fixed seed.
$S^{zz}(k,\omega)$ (Fig.~\ref{fig:szz}) is obtained from a second Lanczos run started on $\sigma_z(k)|0\rangle$; it is quadratic in $\sigma_z$ and so independent of the sign convention.
Its continued fraction is evaluated by Gauss quadrature, which gives pole positions and weights; the weight of the lowest pole is $Z_k$.
At $\gamma=2\Ek$ the quasiparticle peak converges to the infinite-chain dispersion exponentially in $N$; the remainder of the spectral weight lies in a three-particle continuum above $3\Delta$, and the correlation length is below one cell.
At $\gamma=0.6\,\Ek$ ($N=20$) the quasiparticle carries \SIrange{95.6}{99.3}{\percent} of the weight at each $k$ and \SI{97.9}{\percent} integrated over $k$, and the convergence in $N$ is slower; the rest of the weight forms a three-particle continuum that overlaps the band.
The free-fermion references are the finite-$N$ Jordan--Wigner energies in the appropriate parity sector.
The ED energies and poles match the free-fermion values, the first-moment sum rule holds, and the exact static susceptibility $\chi_{j0}(\omega\to0)$ decays with $\kappa a=\ln(2\gamma/\Ek)$.
At $\gamma=\Ek/2$ the gap closes as $N\Delta\to0.786\,\Ek$, close to the conformal value $2\pi\hbar vx_\sigma/a=(\pi/4)\Ek$ with $x_\sigma=1/8$ and $v=a\Ek/\hbar$~\cite{cardy1984conformal}; at the ICHA critical point $\gamma=\Ek$ the gap converges to $\Ek$, so that point is not critical.
Real-time evolution (Fig.~\ref{fig:wavefront}) uses a Chebyshev propagator, which agrees with \texttt{expm\_multiply}, conserves the norm and reproduces the $\mathbb Z_2$ parity of the response to better than $10^{-13}$.
The infinite-chain single-quasiparticle response agrees with ED ($N=20$) to \SI{1}{\percent} ($\gamma=2\Ek$) and \SI{6}{\percent} ($\gamma=0.6\,\Ek$) of the initial tilt for $t\le8\,\hbar/\Ek$; ED reaches only $j\le9$, where both give an apparent front speed of 1.14--$1.20\,a\Ek/\hbar$, a short-distance bias.

\emph{Front speed.}
We locate the front at cell $j$ by the first time $|P_j|$ reaches 1 or \SI{3}{\percent} of the initial tilt in the single-quasiparticle (linear) response of the infinite chain, built from Eq.~\eqref{eq:tfim_exact}, and fit cells $j=20$--120 with the Airy-corrected form $t=j/v+bj^{1/3}+t_0$.
For $\gamma=2\Ek$ this gives $v=0.99\,a\Ek/\hbar$, and 0.97--$1.07\,a\Ek/\hbar$ at $\gamma=0.6\,\Ek$.
The same fits give $1.03\,a\Ek/\hbar$ for the linearized ICHA chain, whose analytic value is $1.035\,a\Ek/\hbar$, and the nonlinear ICHA chain gives $1.02\,a\Ek/\hbar$ (cells 10--65).

\section{Numerical methods for the coherence-vector chain}
\label{app:chain}

The nonlinear chain, Eq.~\eqref{eq:coherence_eom} with nearest-neighbour coupling, is integrated with a fixed-step fourth-order Runge--Kutta scheme in units $\hbar=\Ek=a=1$.
Results for $\diff t=0.01$ and 0.005 differ by less than $10^{-6}$, and for small tilts the nonlinear chain matches the linear lattice sum to $4\times10^{-4}$.
For the damping check [Fig.~\ref{fig:damping}(a)], a 400-cell chain is driven continuously at the carrier on cell 0.
The attenuation is fitted to the steady-state amplitude away from both ends, and it matches the analytic complex root of the linearized chain to better than \SI{0.1}{\percent} for both relaxation models.
With $\tau=40\,\hbar/\Ek$, relaxation toward the instantaneous equilibrium gives $\alpha=0.0236$ per cell, \SI{18}{\percent} below the value $1/(\vg\tau)=0.0288$ per cell obtained when $P$ and $\lambda_y$ relax toward zero.

\section{Junctions: further results and checks}
\label{app:junction}

\emph{Continuum limit of a junction.}
A junction with $\Sigma=2+ga$ becomes, in the continuum limit, a Klein--Gordon node with a point shunt $g$, for which $r=(2\mathrm iK\Gamma_R-g)/(2\mathrm iK+g)$ at wavenumber $K$.
The ordinary junction has $g=(n-1)/a$, which diverges in this limit, so the TL value $\Gamma_R$ is not the long-wavelength limit of a lattice junction.
Near the band edges $|t|^2\simeq4\sin^2ka/(n-1)^2$.
Averaged uniformly over $ka$, $|r|^2=(n-1)/(n+1)$: $1/2$ for a cross (twice the TL power) and $1/3$ for a T (three times); for a cross $|r|^2$ stays within \SI{20}{\percent} of $\Gamma_R^2$ only for $0.34<ka/\pi<0.66$, and for $\gamma=2\Ek$ it reaches 0.63 at $1.05\,\fmin$.

\emph{Junction-bound modes.}
For $\Sigma>2$, $G$ has poles at $e^{2\mathrm ika}=1/(\Sigma-1)$, i.e.\ at $k=\mathrm i\kappa$ and $k=\pi+\mathrm i\kappa$ with $\cosh\kappa a=\Sigma/2\sqrt{\Sigma-1}$.
These are two modes localized on the junction, one below the cutoff and one above the band top:
\begin{equation}
  \hbar^2\omega_b^2=4\gamma\bigl(\gamma\mp\Ek\cosh\kappa a\bigr).
  \label{eq:bound}
\end{equation}
For $\gamma=2\Ek$ they lie at $2.741$ and $4.948\,\Ek$ for a T and at $2.600$ and $5.024\,\Ek$ for a cross, against a band of $2.828$--$4.899\,\Ek$.
The unpolarized state of a junction is stable in the ICHA only for $\gamma>\Ek\cosh\kappa a$, i.e.\ $1.06\,\Ek$ for a T and $1.15\,\Ek$ for a cross.

\emph{ICHA simulation.}
We integrated the nonlinear coherence-vector equations, Eq.~\eqref{eq:coherence_eom}, on star graphs with 450-cell arms and graded absorbing ends.
A 17-tone drive at $ka/\pi=0.10,0.15,\dots,0.90$, with polarizations $|P|\lesssim0.02$, was applied, and two adjacent probe cells per arm separate the incoming and outgoing waves at each tone.
We checked eight configurations: the wire, T and cross; a T with junction bonds $0.7\,\Ek$; the matched cross; a cross with alternating bonds $\Ek$ and $0.6\,\Ek$; and two weak links ($\eta=0.3$, 0.6).
In all of them, $|r|^2$ and $|t|^2$ agree with Eqs.~\eqref{eq:smatrix}, \eqref{eq:r_uniform} and~\eqref{eq:defect} to $4\times10^{-4}$, flux is conserved to $4\times10^{-4}$, and the complex amplitudes agree to $4.3\times10^{-3}$.
A broadband pulse gives the same curves to $4\times10^{-3}$ for $0.2<ka/\pi<0.8$.

\emph{Exact diagonalization.}
We diagonalized Eq.~\eqref{eq:tfim}, extended to the bonds of the graph, for a T with arm lengths $L=5$ and 6 ($N=16$ and 19) and a cross with $L=4$ and 5 ($N=17$ and 21) at $\gamma=2\Ek$.
The one-quasiparticle levels in the arm-symmetric and arm-antisymmetric channels follow from Lanczos runs in the odd-parity sector.
Each channel phase shift $\Phi_{\mathrm{ch}}$ then follows from the finite-size condition $e^{2\mathrm i\Phi_{\mathrm{ch}}}R_{\mathrm{end}}(k)\,e^{2\mathrm ik(L+1)a}=1$, where $k$ comes from the exact dispersion, Eq.~\eqref{eq:tfim_exact}, and $R_{\mathrm{end}}$, the reflection at the free end of an arm, from the exact open-chain solution.
The procedure returns Eq.~\eqref{eq:smatrix} to $10^{-11}$ when applied to the ICHA fan-out, and $|r|^2<10^{-13}$ for an exact straight chain.
In the exact model the antisymmetric channels have $e^{2\mathrm i\Phi_{\mathrm{ch}}}=-1$ to within \SI{4e-3}{rad}, as in the ICHA, where the junction cell is a node.
The reflected power agrees with Eq.~\eqref{eq:r_uniform} at the same $ka$ to within $0.051$ (at most \SI{16}{\percent} relative), and to within $2\times10^{-3}$ at $ka=0.496\pi$; the exact values lie below the ICHA ones in the lower half of the band and above them in the upper half.
The junction binds a quasiparticle below and above the band, as Eq.~\eqref{eq:bound} predicts; the exact binding energy below the band is 0.64--0.66 times the ICHA value at the same $L$: $0.052$ against $0.080\,\Ek$ for the T and $0.149$ against $0.226\,\Ek$ for the cross ($L=5$).
At the junction cell the one-quasiparticle poles carry \SIrange{99.96}{99.98}{\percent} of the $\sigma_z$ spectral weight.

\clearpage
\onecolumngrid
\section{Notation}
\label{app:notation}

Table~\ref{tab:notation} lists the symbols used in the main text and the appendices.

\vspace{-6pt}
\begin{table*}[h!]
\caption{Notation. Energies are in meV unless stated otherwise; a dash marks a dimensionless quantity. ICHA: intercellular Hartree approximation; ED: exact diagonalization; TFIM: transverse-field Ising model; TL: transmission line; JW: Jordan--Wigner.}
\label{tab:notation}
\footnotesize
\renewcommand{\arraystretch}{0.94}
\begin{ruledtabular}
\begin{tabular}{l>{\raggedright\arraybackslash}p{5.0cm}l@{\hspace{1.2em}}l>{\raggedright\arraybackslash}p{5.0cm}l}
Symbol & Meaning & Unit & Symbol & Meaning & Unit \\
\hline
$a$ & cell pitch & nm & $\meff$ & effective rest mass $\hbar\wmin/c^2$ & kg \\
$|{\pm1}\rangle$, $Q_d$ & basis states ($P=\pm1$); charge on dot $d$ & --, $e$ & $\beta$, $\alpha$ & damping rate; attenuation $\beta/\vg$ & s$^{-1}$, nm$^{-1}$ \\
$\sigma_{x,y,z}$ & Pauli matrices, Eq.~\eqref{eq:pauli} & -- & $\ldecay$, $N_c$ & decay length $1/\alpha$; carrier periods within $\tau_{\mathrm{th}}$ & nm, -- \\
$P_i$ & polarization of cell $i$, $-\langle\sigma_z\rangle$ & -- & $\varepsilon_k$, $c_k$ & exact quasiparticle energy, Eq.~\eqref{eq:tfim_exact}; JW fermion & meV, -- \\
$\vlam_i$ & coherence vector, $\lambda_a=\Tr(\rho\sigma_a)$ & -- & $\Delta$, $f_\Delta$ & exact gap $2|\gamma-\Ek/2|$; gap frequency $\Delta/h$ & meV, THz \\
$\gamma$ & intracell tunnelling energy & meV & $\gamma_c$ & critical tunnelling, $\Ek/2$ (ICHA: $\Ek$) & meV \\
$\Ek$ & kink energy & meV & $S^{zz}$, $S(k)$, $Z_k$ & dynamical, static structure factor; pole weight & meV$^{-1}$, -- \\
$\Eeff$ & field of the neighbours, $\sum_jE_{k,ij}P_j$ & meV & $\tau_{\mathrm{th}}$, $x_\sigma$ & thermal dephasing time; scaling dimension of $\sigma_z$ & ps, -- \\
$\vec\Gamma_i$ & $\Gamma_a=\Tr(\sigma_aH_i)$, Eq.~\eqref{eq:gamma_vec} & meV & $n$, $\eta_\alpha$, $\Sigma$ & output arms; bond of arm $\alpha$ over $\Ek$; $\sum_\alpha\eta_\alpha^2$ & -- \\
$\tau$ & relaxation time & ps & $\mathcal S_{\alpha\beta}$, $G$ & junction scattering matrix; junction function & -- \\
$T$, $\theta$ & temperature; $\tanh(\gamma/\kB T)$ & K, -- & $r$, $t$, $\Gamma_R$ & reflection; transmission per arm; TL reflection & -- \\
$k$, $\kappa$ & wavenumber; decay constant below cutoff & nm$^{-1}$ & $\omega_b$, $g$ & junction-bound mode; continuum point shunt & s$^{-1}$, nm$^{-1}$ \\
$\omega$, $f$ & angular and ordinary frequency & s$^{-1}$, Hz & $\eta$ & weak-link bond (over $\Ek$), Eq.~\eqref{eq:defect} & -- \\
$\wmin$, $\fmin$ & ICHA cutoff, Eq.~\eqref{eq:fmin} & s$^{-1}$, Hz & $f_c$ & reference carrier, $\simeq1.13\,\fmin$ & THz \\
$\omega_0$, $\Omega_I$ & Telegrapher frequency; ICHA growth rate & s$^{-1}$ & $T_{\max}$ & temperature ceiling, Eq.~\eqref{eq:tmax} & K \\
$c$ & Klein--Gordon velocity $a\sqrt{2\gamma\Ek}/\hbar$ & km/s & $\phi_0$, $\Phi_{\mathrm{ch}}$ & initial tilt (Fig.~\ref{fig:wavefront}); channel phase shift & rad \\
$\vp$, $\vg$ & phase and group velocity & km/s & $W$, $s_\omega$, $\mu_k$ & Lambert function; rate and coupling ratio, Eq.~\eqref{eq:charpoly} & --, s$^{-1}$, -- \\
$\chi_{j0}$ & exact susceptibility of cell $j$ to a field on cell 0 [Fig.~\ref{fig:damping}(b)] & meV$^{-1}$ & $J$ & Ising exchange $\Ek/2$, Eq.~\eqref{eq:tfim} & meV \\
\end{tabular}
\end{ruledtabular}
\end{table*}
\twocolumngrid

\bibliography{references}

\end{document}